\documentclass[12pt,preprint]{aastex}

\begin{document}

\shorttitle{}
\shortauthors{Herbst  \& Mundt}



\title{Rotational Evolution of Solar-Like Stars in Clusters from Pre-Main Sequence to Main Sequence: Empirical Results}


\author{William Herbst\altaffilmark{1}}
\affil{Astronomy Department, Wesleyan University, Middletown, CT
06459}

\email{wherbst@wesleyan.edu}

\and

\author{Reinhard Mundt}
\affil{Max-Planck-Institute for Astronomy - Heidelberg, Ko\"nigstuhl
D-17, Germany}


\altaffiltext{1}{Visiting Scientist at MPIA-Heidelberg}


\begin{abstract}
Rotation periods are now available for $\sim$500 pre-main sequence
and recently arrived main sequence stars of solar-like mass (0.4-1.2
M${_\odot}$) in five nearby young clusters: the Orion Nebula Cluster, NGC 2264, {$\alpha$} Per, IC 2602 and the Pleiades. In combination with
estimates of stellar radii these data allow us to construct
distributions of surface angular momentum per unit mass at three
different epochs: nominally, 1, 2 and 50 My. There are sufficient
data that rotational evolution can now be discussed statistically on
the basis of the evolution of these distributions, not just on the
evolution of means or ranges, as has been necessary in the past.  Our main result is illustrated in
Fig. \ref{j_hists_uncorrected} and may be summarized as follows: 
(1)  50-60\% of the stars on convective tracks in this mass range are released from any 
locking mechanism very early on and are free to conserve angular
momentum throughout most of their PMS evolution, i.e. to spin up and
account for the rapidly rotating young main sequence stars.
(2) The other 40-50\% lose substantial amounts of angular momentum during the first few million years, and end up as slowly rotating main sequence stars. The duration of the rapid angular momentum loss phase is $\sim$5-6 My, which is roughly consistent with the lifetimes of disks estimated from infrared surveys of young clusters.
The rapid rotators of Orion age lose less than 10\% of their (surface) specific angular momentum during the next 50 My while the slow rotators lose about two-thirds of theirs.  A
detectable part of this loss occurs even during the $\sim$1 My
interval between the ONC and NGC 2264. The data support the view that
interaction between an accretion disk and star is the primary
mechanism for evolving the broad, bimodal distribution of rotation
rates seen for solar-like stars in the ONC into the even broader distributions seen in the young MS clusters.
\end{abstract}




\section{Introduction}

The early evolution of rotation of solar-like stars (0.4-1.2
M$_\odot$ for the purposes of this paper) is a fundamental and
surprisingly controversial subject that has recently attracted a good
deal of theoretical and observational attention. For reviews of some
recent conference discussions see \citet{m03} and \citet{st03}. A
central issue has been to understand how the observed broad range of
rotation rates, which is about a factor of 20 for PMS stars and
larger for recently arrived MS stars, comes to exist and how it
evolves. The physical phenomena that have been proposed, modeled and
debated include the overall contraction of the star during its PMS
phase with expected conservation of angular momentum, accretion which
can either add or subtract angular momentum per unit mass,
magnetically driven stellar winds which drain angular momentum, and
internal redistribution of angular momentum. For comprehensive recent
discussions and references to the earlier literature see, for
example, papers by \citet{kpb97}, \citet{spt00} and \citet{b03}. 

An area of particular importance (and controversy) is the putative
role of ``disk-locking", the theory that the angular velocity of the
stellar surface may be locked to a location in the accretion disk
several stellar radii above the photosphere. Originally proposed by
\citet{c90} and \citet{k91} as an explanation for the slow rotation
seen in many classical T Tauri stars (CTTS) it has become an
essential feature of most, if not all, models of angular momentum
evolution of solar-like stars [for example, \citet{bsp01} and
\citet{tpt02}]. At the same time, however, the concept has been
criticized on observational and theoretical grounds, most recently by
\citet{mp04}. As F. Shu discusses in the conference report by
\citet{st03}, part of the problem is that there is no current ``first
principles" model of disk-locking. The physical complexities of a
stellar dynamo linked to an external disk are too much to contend
with at present. Another problem is that the observed correlations
between rotation and accretion disk indicators, although highly
significant, are nonetheless weak in the sense of there being a good
deal of scatter in the data \citep{hbm02, lbm04}. While there are
good reasons for not expecting a tight correlation, including the
difficulties of observing accretion disks with inner holes and the
time lapse between the disappearance of active accretion and the
substantial spin-up of the star due to contraction, doubts are raised
on the observational side by some authors \citep{m03}.

Some recent analyses of the growing data base on stellar rotation
have also led to puzzling and contradictory conclusions with respect
to disk-locking and rotational evolution. \citet{rws02} find that
solar-like stars do not appear to conserve angular momentum as they
contract during their first 3 My of existence, suggesting the
importance of disk-regulation. Yet, they also do not find the
near-infrared excesses expected of disks for most stars. A similar
puzzle is raised by \citet{mrs04} who find that while the mean size
of stars (of similar mass) in the extremely young cluster NGC 2264 is
smaller than in the Orion Nebula Cluster, the period distribution in
NGC 2264 is indistinguishable from that in Orion. This, again,
suggests that most stars contract without spinning up. Again,
however, they cannot relate this to any observational evidence for
the presence of disks around such a large fraction of the stars or
any correlation between presence of a disk and rotation. Finally, in
this vein,  \citet{rws04} find that ``a significant fraction of all
pre-main-sequence (PMS) stars must evolve at nearly constant angular
velocity during the first $\sim$3-5 Myr after they begin their
evolution down the convective tracks." In order to explain the rapid
rotators (also known as ``ultrafast rotators") in older clusters such
as $\alpha$ Per and IC 2602, these authors also argue, however, that
at least 30-40\% of the PMS stars in their sample cannot actually be
regulated for times exceeding 1 My. To summarize this body of work,
Rebull and collaborators interpret the data on rotation periods of
extremely young clusters to indicate that the majority of stars must
be regulated for up to 4 My during their PMS contraction phase. 

A very different picture has recently been proposed by
\citet{lbm04,lmb04}. Based on their more extensive observational
study of rotation in NGC 2264 they conclude that while, indeed, the
NGC 2264 stars are smaller than their counterparts in the ONC, they
also generally spin faster. A detailed description and account of the
differences between the \citet{mrs04} and \citet{lmb04} results is
given in the latter paper and need not be repeated here, but some
aspects of the distinctively different interpretations will be
revisited in what follows. 

The purpose of this paper is to re-analyze the existing data on
stellar rotation in extremely young clusters in an attempt to clarify
exactly what the observations say and do not say about the evolution
of rotation of solar-like stars from the PMS to the MS. This new look
is warranted, we believe, because there are finally enough rotation
periods for stars in the relevant mass range at sufficiently
different ages to allow a statistically valid comparison which
employs the full distribution of rotation periods, not just a median
or range. Section II of the paper describes and justifies our
approach to the subject, which is as empirical and as
model-independent as we can make it. Section III presents the
distributions of the relevant quantities, rotation periods, stellar
radii and specific angular momenta at three different epochs
(nominally 1, 2 and 50 My). In Section IV we discuss the results in
terms of other current work and Section V summarizes our conclusions.

\section{Analysis}

\subsection{Basic Issues and Assumptions}

The magnitude of the angular momentum vector of an object rotating
about a spin axis is $$J = I \omega$$
where I is the moment of inertia and $\omega$ is related to the
rotation period (P) by $$\omega = {{2 \pi} \over P}.$$ The moment of
inertia may be expressed in terms of the ``radius of gyration", kR,
for a spherical body of radius R. The radius of gyration is the
distance from the spin axis that one would have to place a point
mass, equal to the object's mass (M), to obtain the equivalent moment
of inertia. Hence, $$I = M (kR)^2.$$ The value of k depends on the
mass distribution within the object, as well as its shape. Since
rotating stars become more and more distended at low latitudes with
increasing spin rate,  k is a function of P. Combining the equations
above, one can write that the specific angular momentum, j, is given
by $$j = {J \over M} = {{{2 \pi k^2}{R^2 \over P}}}.$$ It is the
evolution of the specific angular momentum with time that we seek to
constrain and it is simply related to only three variable quantities,
rotation period and radius, both of which may, at least in principle,
be determined from the observations and radius of gyration, which can
be approximated with good accuracy from a theory of rotating
polytropes \citep{c35,j64}. 

The particular approach to studying the evolution of j adopted here,
which differs in important ways from other approaches in the
literature, is motivated by certain characteristics of the
observations as well as by the fact that R evolves rapidly during PMS
contraction. We note that rotation periods have a range of 20 or
more, that their distribution is highly mass dependent, but that they
can be measured to a high degree of accuracy ($\sim$1\%). Radii, on
the other hand are poorly determined for any one star but expected to
have a vanishingly small range for stars of the same mass and age and
a relatively weak mass dependence. Radii of 1 My old solar-like stars
are expected to evolve quickly and, for angular momentum
conservation, P $\propto$ R$^2$, so P may be very sensitive to the
precise age of the star. Therefore, to best constrain the evolution
of j with time, one wishes to have a large sample of stars within a
narrow mass range and with very nearly the same ages. This will allow
one to accurately define the broad period distribution characteristic
of a particular time and to allow the large scatter in measured radii
to be dealt with by averaging. Clearly, what one requires is populous
clusters, where there are a good number of stars of appropriate mass
whose coevality is reasonably guaranteed by their cluster membership.
 The use of clusters in this way is nothing new, of course --- it is
simply the {\it classical} method used to investigate all aspects of
stellar evolution, which has been employed by astronomers for more
than a century. 

Another element of our approach is that we employ only rotation
periods in establishing the angular momentum distribution, not v sin
i measurements. The reason is that rotation periods are known for a
sufficient number of stars in the relevant mass range (0.4-1.2
M$_\odot$) and that there is no need to introduce the complications
that arise from v sin i studies. These include the inherent
uncertainty due to the unknown inclination of the system, and the
problem of accuracy, especially for slowly rotating stars. v sin i
measurements are very helpful at verifying rotation periods and
resolving issues of harmonics or beat periods as discussed below.
They are also useful for determining whether selection effects are
present in the rotation properties determined from the periodically
variable stars. They do not, however, contribute in an important way
to the definition of the angular momentum distribution in a cluster
when numerous direct and accurate measurements of $\omega$ are
available. Hence, we focus only on rotation periods. An implicit
assumption in doing this is that the subset of cluster stars with
known periods is unbiased with respect to rotation. In fact, this is
unlikely to be true for all clusters and we return to the issue in
Section IV. 

Finally, we mention a perhaps obvious point that we wish to make
explicit. Our analysis initially assumes that there are not
cluster-to-cluster differences of significance in the initial
rotation period distribution. In other words, we take all differences
to be the result of evolution from a common starting distribution
represented here by the ONC. No progress can be made empirically
without this assumption since there is no theory or set of
observations that can yet tell us how the P distribution at 1 My is
created or how much variance in that we should expect between
clusters. Hence, we proceed with the assumption of a common starting
P distribution and, at the end of the analysis, consider whether any
element of the results points to such cluster-to-cluster differences.
Since every cluster must evolve under at least slightly different
environmental conditions it is not hard to imagine ways in which the
initial angular momentum distributions could differ. Perhaps the
surprising thing is that we ultimately find only weak evidence for
some small differences in starting P distributions.

\subsection{Sample Selection: Clusters and Mass Range}

The first step in our procedure is to identify clusters at a range of
ages that have sufficient rotation results to be useful. There are
two rich PMS clusters which are perfect for this work --- the ONC and
NGC 2264. Over 400 rotation periods are known in the ONC from the
studies of \citet{mh91}, \citet{ah92}, \citet{ehh95}, \citet{ch96},
\citet{smm99}, \citet{hrh00}, \citet{chs01} and \citet{hbm02}. A
similar number are known in NGC 2264 from work by \citet{keh97},
\citet{kh98}, \citet{lbm04}, \citet{mrs04} and \citet{lmb04}. Other
small clusters such as IC 348 and NGC 1333 are not included because
they add very few stars in the relevant mass range. Association and
field stars are not included for the reasons discussed in the
previous section. In particular, we emphasize the distinction that
exists between the Orion association (Ori OB I) and the ONC. It has
been known since the work of \citet{b64} and \citet{wh78} that the
Orion star forming region is a complex one with a variety of
sub-associations of different ages and one extremely populous, dense
cluster (the ONC). The definition of the ONC that we employ is the
one used by \citet{h97}. In particular, it does not include the
regions studied by \citet{r01}, which she calls the Orion ``flanking
fields" or the survey (apart from any ONC stars) of \citet{chs01}
which covers a wide range of Orion association stars. This
distinction between the ONC and the ``Orion region" is central to our
analysis and further sets it apart from the recent discussion by
\citet{mrs04}. For additional discussion of this critical point see
\citet{lmb04}. 

The other three clusters which we employ are the Pleiades, $\alpha$
Per and IC 2602. These were, again, selected because there are
extensive photometric surveys which have found rotation periods for
dozens of stars. Their ages are much greater than the ONC and NGC
2264 and usually quoted to be in the range 50 to 120 My. The
solar-like stars in these clusters are on or close to the MS so the
rapid phase of evolution of radius is completed and we will refer to
them as the ``MS clusters", to distinguish them from the ONC and NGC
2264. The period and angular momentum distributions of the MS
clusters are much more similar to each other than any of them are to
the ONC or to NGC 2264. As we show in what follows, the rather small
differences in rotation properties that do exist among them can be
understood in terms of the age range, wind losses and perhaps
selection effects. Since the primary focus of this paper is the much
more dramatic angular momentum losses that accompany the evolution of
some stars from PMS to MS, we will combine the data on the MS
clusters, correcting for the small differences probably caused by
wind losses or selection effects, to create a substantial set of
stars defining the rotation properties of recently arrived or soon to
arrive MS stars. The authors express here their gratitude to J.
Stauffer for maintaining the excellent data base on young clusters
assembled by the late Charles Prosser, on which the present results
are based.

Our selection of the mass range to study is dictated by the
availability of rotation periods within the MS clusters. Currently,
there is good coverage available only for stars with effective
temperatures between log T$_{eff}$ of about 3.80 and 3.55
corresponding to masses between about 1.2 and 0.4 M$_{\odot}$. It is
increasingly difficult to get rotation periods for lower mass stars
because they are faint and red, increasing the noise while reducing
the signal as the spot-photosphere contrast decreases. In order that
we compare apples with apples it is necessary to know what range of
log T$_{eff}$ of PMS stars corresponds to this mass range on the MS.
Unfortunately, there is no agreement yet among PMS modelers on
precise evolutionary tracks; the situation is nicely exhibited on
Fig. 1 of \citet{hw04}. Fortunately, however, as one can see from
that figure, regardless of the particular tracks followed the models
do at least agree on the range of log T$_{eff}$'s on the PMS that
will map onto the range of MS log T$_{eff}$'s over which periodic
stars are actually measured. This is approximately the range log
T$_{eff}$= 3.54-3.67, corresponding to spectral classes of K4 to M2
if one adopts the \citet{ck79} calibration (see below). Many stars
with this T$_{eff}$ range in the ONC and NGC 2264 should end up
as MS objects with masses between 0.4-1.2 M$_\odot$ regardless of the
precise tracks followed. We are aware that on the high and low-mass
ends of the selected PMS star log T$_{eff}$ range some stars may end
up on the main sequence outside of the considered mass range, but due
to the weak dependence of j on mass (see below) this will not
influence our conclusions significantly.

\begin{figure}
\plotone{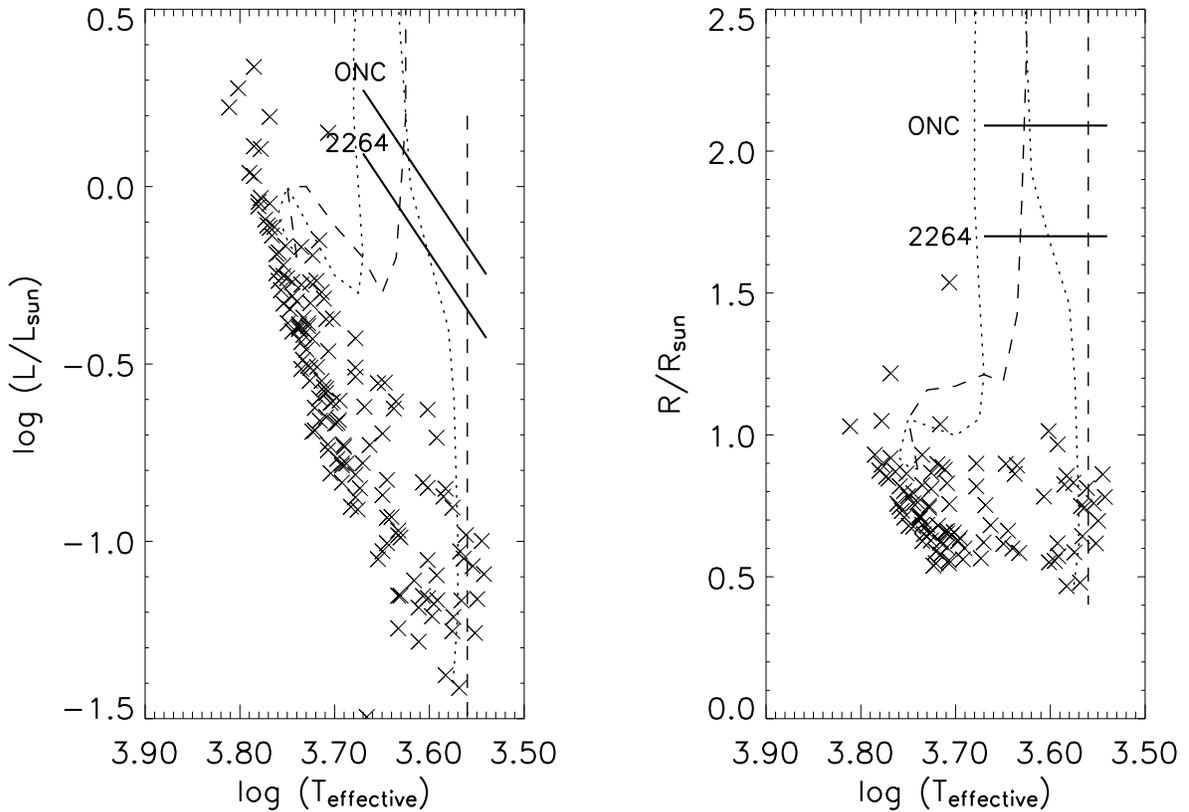}
\caption{The adopted
cluster loci for the ONC (age $\sim$1 My) and NGC 2264 (age $\sim$2 My) are shown as solid lines on
this theoretical HR diagram (left) and on a plot of radius versus log
T$_{eff}$ (right). Individual stars in the three older clusters ($\alpha$ Per, IC 2602 and the Pleiades) are
plotted. Also shown are theoretical tracks for 1 and 0.5 solar mass
stars as calculated by \citet{dm97} (dotted line) and for 1 and 0.4
solar masss stars by \citet{ps99} (dashed line). These more or less
bracket the range of paths shown by such models \citep{hw04}.}
\label{hr_diagram}
\end{figure}

Fig. \ref{hr_diagram} shows the loci of stars selected for this study
on an HR diagram. The ONC and NGC 2264 are represented by mean
relations based on the average radius employed (see below).
Individual stars are plotted for the MS clusters. Overlain on this
are two sets of theoretical tracks which illustrate the range of
results obtained by modelers. Tracks for a 1.0 and 0.5 solar mass
star are shown from \citet{dm97} and tracks for a 1.0 and 0.4 solar
mass star from \citet{ps99}. It is evident that, depending on whose
results are adopted, one could predict a MS mass that differed by a
factor of 2 for a star with given values of luminosity and effective
temperature. The real problem is compounded, of course, by the
difficulties of  determining accurate values of luminosity and
effective temperature for PMS stars (see below) and by the possibly
important factors such as rotation and magnetic fields which are
neglected in all of the models. For an evaluation of the potential
importance of magnetic effects see \citet{dvm00}. However, it turns
out that the models do at least agree on the point that the range of
log T$_{eff}$ among PMS stars that will map onto the MS in the 0.4 -
1.2 M$_{\odot}$ range is the adopted range of 3.54 to 3.67,
corresponding to a spectral class range of K4 to M2.  

Fortunately, neither the MS nor the PMS rotation data suggests that j
is a strong function of mass, so that if we have mis-matched the mass
ranges in the PMS and MS clusters to some degree it should not have
an important impact on the results. \citet{hbm01} reviewed the
situation in the ONC and showed that, while rotation period, indeed,
has a clear dependence on mass, j varies little, if at all. In other
words, while lower mass stars on the PMS do rotate significantly
faster than their higher mass counterparts, they are also smaller by
enough to leave j nearly a constant across masses. Also, the
differences among PMS stars do not become obvious until a mass lower
than those considered here, namely $\sim$0.25 M$_{\odot}$ on the
\citet{dm97} scale, is reached. We show in what follows that j is
also independent or nearly independent of mass for MS stars in the
relevant mass range. This circumstance relieves some of the pressure
to be certain that the range of PMS stars selected is precisely the
one that will map onto the MS at 0.4-1.2 M$_{\odot}$.  PMS stars of
all masses, within the 0.08-2.0 M$_{\odot}$ range where rotation
period data is available, have about the same range and distribution
of j. 

\subsection{Rotation Periods}

This work is based on the assumption that the photometric periods
derived for G,K and M stars in young clusters result from the
rotation of a spotted surface and that the photometric period is, at
least in most cases, an accurate reflection of the stellar surface
rotation period. While there is little or no controversy on these
points in the literature it is, perhaps, worth briefly reviewing the
evidence for this assertion. First, we note that photometric periods
can be reliably and accurately determined. A five-year study of the
young cluster IC 348 by \citet{chw04} illustrates that, while there
are changes in the light curve shapes from year to year, and that in
some seasons the light variations become incoherent, when a period is
found it is always the same period to within the errors, which are
typically 1\%. Early reports of significant changes in periods for
the same star \citep{bcf93} have not been confirmed nor have
additional cases been reported. It is most likely a result of
applying an inappropriate false alarm probability to the data which
did not reflect the correlations that exist in the photometry
\citep{hw96, r01}. 

The shapes of the light curves, the periods involved, the color
behavior, the evolution of light curve forms, the amplitudes and the
correlation of period with v sin i measurements all support the
identification of the photometric period with the rotation period of
the star and the source of the variations as spots (primarily cool,
but sometimes hot) on the stellar photospheres. For the PMS stars,
the photometric amplitudes indicate enormous spots covering
substantial portions of the star's surfaces and most likely situated
at high latitudes. So far, there has been no definitive determination
of any period change with time that would indicate spot migration and
differential rotation as is seen, for example, on the sun. The
periods repeat to within their errors for all stars observed in
multiple seasons and by multiple observers, except as noted already.
This gives us confidence that the photometric period is a measurement
of a fundamental stellar property, namely the surface rotation rate.
For the MS stars, the situation is similar except the amplitudes of
variation tend to be much smaller -- only a few percent, at most.
Again, the fact that the periodicity is measuring stellar rotation
rate of stars in the MS clusters is affirmed by the excellent
correlation of P with v sin i. 

When different investigators, using different telescopes, observing
procedures, wavelengths, period search algorithms and false alarm
indicators have studied the same clusters they have found very
compatible results. For the ONC, a comparison between the periods
determined by \citet{smm99} and by \citet{hbm02} has been given in
the latter work and it shows agreement to within the errors for 85\%
of the 113 stars in common between the studies. Cases of disagreement
are almost always the result of harmonics (i.e. sometimes a star can
develop spots on opposite hemispheres so that the rotation period is
actually twice the photometric period) or beat periods with the
observing frequency of once per night. Comparison of data sets
obtained at different epochs and with different observing frequencies
helps eliminate these cases, as does comparison with spectroscopic v
sin i measurements \citep{rhm01}. Very similar results were obtained
in comparing periods in NGC 2264 reported by \citet{mrs04} with those
found by \citet{lmb04}, even though the epochs of observation were
years apart. Again, there were 113 stars in common between the
studies and agreement to within the errors was found for all but 15
of them, i.e. 87\% agreement. See \citet{lmb04} for further
discussion of this comparison.
 
One final comment about rotation periods is, perhaps, in order. It is
good to keep in mind that we measure only the {\it surface} rotation
rate of the stars. Throughout this analysis, when we speak of
rotation rate, that is what we mean. There is no guarantee and,
indeed, no current way of knowing how the rotation rate varies with
depth in the star. It is plausible that PMS stars rotate nearly as
rigid bodies because they are believed to be fully convective. Hence,
mass and angular momentum may be efficiently mixed from the surface
to the core. On the MS, stars in this mass range have radiative
cores, so it is unlikely that they are similarly mixed. From the
observational perspective, it is actually impossible to say and one may consider limiting cases such as conservation in shells or solid body rotation \citep{wsh04}. It is certainly true that the angular momentum loss implied by our data is
substantial for some stars and it would clearly be much easier for a star to lose
such angular momentum from a relatively narrow surface shell than
from its entire mass. So, we caution the reader once more that the
only quantity that can be observed is the angular momentum per unit
mass that applies to the surface shell, not an average j appropriate
to the whole star. What the profile of rotation with depth is in any
star other than the Sun is currently a question that can be addressed
only by theory.

\subsection{Radii}

Besides rotation periods, which are known to an accuracy of
$\sim$1\%, one requires only stellar radii to determine j for a
spherical star. Unfortunately, radii are very poorly known for
individual PMS stars. The reason is that they can currently only be
calculated from the fundamental relationship $L = 4 \pi R^2 \sigma
{T_{eff}}^4$, which requires that luminosity (L) and T$_{eff}$ be
determined. L, in turn,  depends on the apparent brightness (which
fluctuates nightly and even hourly for essentially all PMS stars), an
extinction correction (which not only depends on establishing the
intrinsic color and color excess but also on an assumed reddening law
that could be abnormal in star forming regions such as Orion), a
bolmetric correction and a distance. Fortunately, the radius does
only depend on the square root of the luminosity and is, therefore,
linearly dependent on the distance.

Perhaps the largest uncertainty in radius determination for PMS stars
is the translation required between an observable quantity, spectral
type (or, class, actually), and the theoretical quantity T$_{eff}$.
It is instructive to recall that the relationship which is still most
commonly employed in the mass range of interest here, is the one
proposed by \citet{ck79}. While one might hope that the longevity of
this relationship is due to the rigor with which it was assembled,
the authors' own comments dispel any such notion. They clearly
regarded the relationship they constructed in the 1970's as uncertain
and subject to revision. They furthermore commented on the
fundamental difficulty of ever obtaining an accurate T$_{eff}$
measurement for a PMS star, which may be summarized in this way. The
atmosphere of a T Tauri star is highly magnetized and, as a result
heterogeneous in terms of temperature. If it were not, we would not
be able to detect rotation periods photometrically. On the other
hand, a basic assumption of the expression $L = 4 \pi R^2 \sigma
{T_{eff}}^4$, is that the radiation from the star is isotropic, so
that it can be determined by sampling what is received in the tiny
solid angle defined by the Earth at the distance of the star.
Clearly, this is not true, so fundamentally there is a difficulty in
ever calculating accurate values of T$_{eff}$ for such heterogeneous
atmospheres. Besides this fundamental problem, there are a large
number of practical difficulties in computing L, T$_{eff}$ and
therefore R, which do not need to be reviewed here. The interested
reader is referred to \citet{h03} and \citet{rws04} for further
discussion. Following \citet{h97} we adopt the attitude that there
is, unfortunately, little more that can be done to improve the
situation over what \citet{ck79} did and we will simply adopt the
radii calculated by her for the ONC stars and by \citet{rws02} for
the NGC 2264 stars, which is based on the \citet{h97} approach. 

One difference between our analysis and others is that we do not rely
on individual stellar radii to establish the age of a particular star
or determine whether it belongs to a cluster. There is a huge scatter
in the radii of stars in the ONC which, if interpreted literally,
would mean that star formation has been ongoing there for about 10
My. While some authors have interpreted the scatter to indicate
precisely that \citep{ps00}, we subscribe to the view argued forcibly
by \citet{h03} that, in fact, star formation in the ONC was a rapid
process and that the large majority of the stars seen projected on
the cluster have a single, common age of $\sim$1 My, Hence, we use
membership in the cluster (which is based almost exclusively on
location on the sky) as the primary source of identifying a set of
stars of homogeneous age and regard the large scatter in calculated
radii as simply indicative of the difficulties of measuring the
quantity. We show in what follows that, by using an average radius
versus spectral class relationship in place of individual radii we
get a somewhat tighter j distribution in the ONC.

\begin{figure}
\plotone{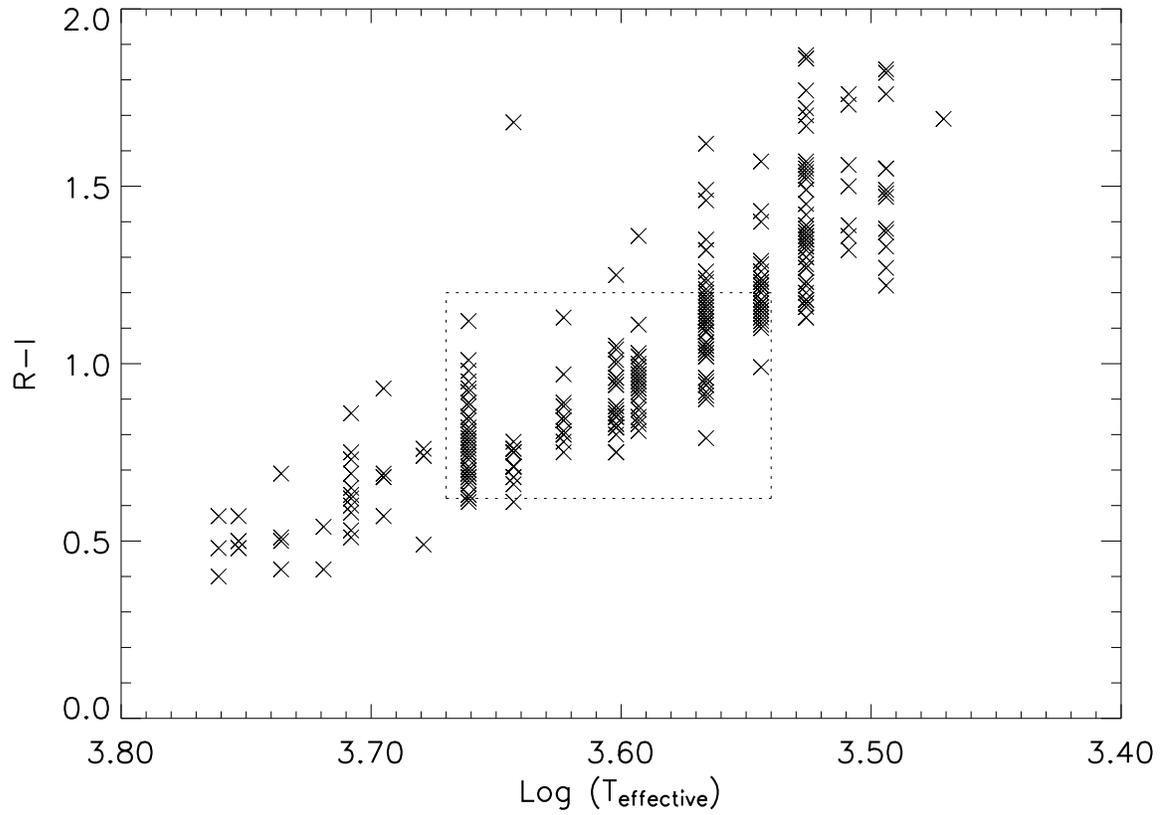}
\caption{R-I versus log
(T$_{eff}$) for stars with known spectral types in NGC 2264. Colors
are from \citet{lbm04} or from \citet{rws02}, spectral types are from
\citet{rws02} and the calibration of log (T$_{eff}$) with spectral
type is from \citet{ck79}. It may be seen that by choosing R-I values
between 0.62 and 1.20 we will select a sample dominated by stars
within the desired log (T$_{eff}$) range of 3.54-3.67.} 
\label{logte_ri_2264}
\end{figure}

While individual radii are available for almost all of the periodic
ONC stars from the extensive study of \citet{h97}, the same is not
true for NGC 2264. Radii are available for 65 of the 182 periodic
stars in the relevant mass range from \citet{rws02}. As noted above,
these were calculated from the spectral types using the same
procedure as \citet{h97} so there should be no systematic errors
involved in their usage other than those which affect the ONC radii.
It is also, fortunately, possible to use color as a surrogate
measurement of log T$_{eff}$ in NGC 2264 because the reddening in the
cluster is known to be fairly small and rather uniform, at least by
comparison with the ONC \citep{rws02}. One may, therefore, expect a
reasonable correlation between observed color (R-I) and log
T$_{eff}$. This is shown in Fig. \ref{logte_ri_2264} for the stars in
our periodic sample which have radii determined by \citet{rws02}.
Clearly there is a good correlation in spite of the fact that
reddening effects induce some scatter, additional to the usual causes
of such scatter, at each spectral class. Our approach is to limit the
sample studied to R-I values between 0.52 and 1.10 which, as may be
seen, provides a set of stars with about the right logT$_{eff}$
range. Some heavily reddened earlier-type stars or lightly reddened
later-type stars may also be selected but this ``pollution" of the
sample is not important because the numbers are few and because there
is little or no dependence of j on mass. It should be kept in mind
that the scatter in the relationship between color and spectral class
may not be entirely due to reddening effects --- it could also have a
significant component caused by errors in the spectral classification
process.

Radii for the MS cluster stars are, of course, more reliable than for
the PMS stars because the stellar atmospheres are less heterogeneous
(photometric amplitudes due to spots do not exceed a few percent),
the spectral energy distributions are less complex (allowing easier
calibration of bolometric corrections, for example) and there is
relatively large sample of nearby stars to use as calibrators. The
exact procedure used to assign radii is not critical for this study
since the breadth of the period distribution is very large compared
to the range of radii. It is the period distribution that dominates
the form of the angular momentum distribution, not the radii. The
procedure adopted is to use the B-V photometry in Prosser's on-line
catalogue along with the calibrations of log T$_{eff}$ and bolometric
correction of \citet{bcp98} to move to the theoretical plane. Uniform
reddening, a standard extinction law with a ratio of total to
selective extinction of 3.1 and the distance moduli recommended by
Prosser were also adopted. For the Pleiades, these values are: mean
reddening, E(B-V) = 0.04, and distance of 127 pc. For the $\alpha$
Per cluster, the values were E(B-V)=0.1 and a distance of 165 pc. For
IC 2602, we adopted E(B-V)=0.04 mag and a distance of 150 pc. 

\subsection{Binary Stars}

One issue in this analysis is how to handle binary stars. The effect
of binaries is easily seen on the MS cluster HR diagrams because the
single-star locus is so well defined and the errors in the photometry
and calibrations are relatively small compared to the sometimes large
effect that a binary companion can have. A binary sequence parallel
to and above the MS is readily apparent (see Fig. \ref{hr_diagram}). This can increase the
apparent luminosity of a stellar system by up to 0.75 mag which, in
the worst case, could cause an over-estimate of the radius by
$\sqrt{2}$ if the star were mistaken for single. That, in turn, could
cause j to be over-estimated by a factor of 2. Fortunately, even that
sort of effect is relatively small for this analysis because the
range in j within the MS clusters is about a factor of 100.
Furthermore, we have no way at present of identifying the binaries in
the PMS clusters. Hence, our approach has been to ignore them in both
types of clusters because the effects are relatively small and
because we hope they will cancel to a better extent if the binaries
are not removed preferentially from one sample (the MS clusters) but
not the other. For completeness, we illustrate in what follows the
effect of correcting the j-distributions of the MS clusters for the
presence of binaries. For the purpose of that comparison, radii were
calculated from log T$_{eff}$ values alone, assuming a MS relation
fit by eye that ignores the binary sequence. As expected, we find
that the correction for binaries in this manner leads to only a small
difference in the j-distribution, shifting some stars to smaller
values and conclude that binaries may be safely ignored in this
analysis.

\subsection{Cluster Ages}

The precise ages of the five clusters employed here are less
important than their ratios. The adopted age scale is based on two
fiducial points, 1 My for the ONC and 120 My for the Pleiades.	The
ONC age comes from the analysis of \citet{h97} who actually derives
0.8 My as her best estimate, but given the uncertainties in the
models and transformations from observational to theoretical plane
already discussed this is consistent with 1 My. The Pleiades age is
based on discussions in the literature by \citet{ssk98} and
\citet{tsp00}. The age of NGC 2264 follows from the ONC by comparing
the luminosity of PMS stars of the same spectral class. Two recent
studies agree that the cluster is about a factor of two older than
the ONC, which places its age at about 2 My \citep{mrs04,lmb04}. Ages
for the other two MS clusters, $\alpha$ Per and IC 2602 are not that
important for this analysis since we are not concerned with the
relatively small changes in j that occur during the MS phase,
presumably due to wind losses and internal angular momentum
transport. Representative estimates from the literature, based on
Li-depletion, are 50 My for IC 2602 \citep{ran01} and 70-75 My for
$\alpha$ Per \citep{bm99}. It is generally agreed that the age order
from low to high for the MS clusters is IC 2602, $\alpha$ Per, the
Pleiades. 

\subsection{Radii of Gyration}

Finally, we need to consider the radius of gyration, kR, since some
stars in our sample spin fast enough that they must be significantly
distorted from a spherical shape. As noted above, we only consider in
this empirical study, the surface rotation and, therefore, need only
be concerned with the surface shape. Fortunately, the problem of
equilibrium shapes of rotating stars has been solved analytically for
polytropes by \citet{c35}. Here we approximate PMS stars as
polytropes of index n=1.5 and ZAMS stars as polytropes of index
n=3.0. Following \citet{c35} we may then write that the surface of a
rotating star is defined by  
$${R(\theta) \over R_o} = {a - b P_2(\theta)}$$ 
where $\theta$ is the usual polar angle, $R_o$ is the radius of the
non-rotating star, and $P_2(\theta)$ is the second order Legendre
polynomial. For a PMS star (n=1.5) and ZAMS star (n=3) respectively, 
$$ a =	1.74225 v + 1\ (for\ n=1.5), \ \ a =  1.99496 v + 1\ (for\
n=3)$$ and
$$ b = 3.86184 v\ (for\ n=1.5), \ \  b = 27.8734 v\ (for\ n=3)$$
where,
$$v = {\omega^2 \over {2 \pi G \rho_c}}.$$
It is clear that for a given polytropic index the shape depends only
on $\omega$ and the central density of the star ($\rho_c$).

\begin{figure}
\plotone{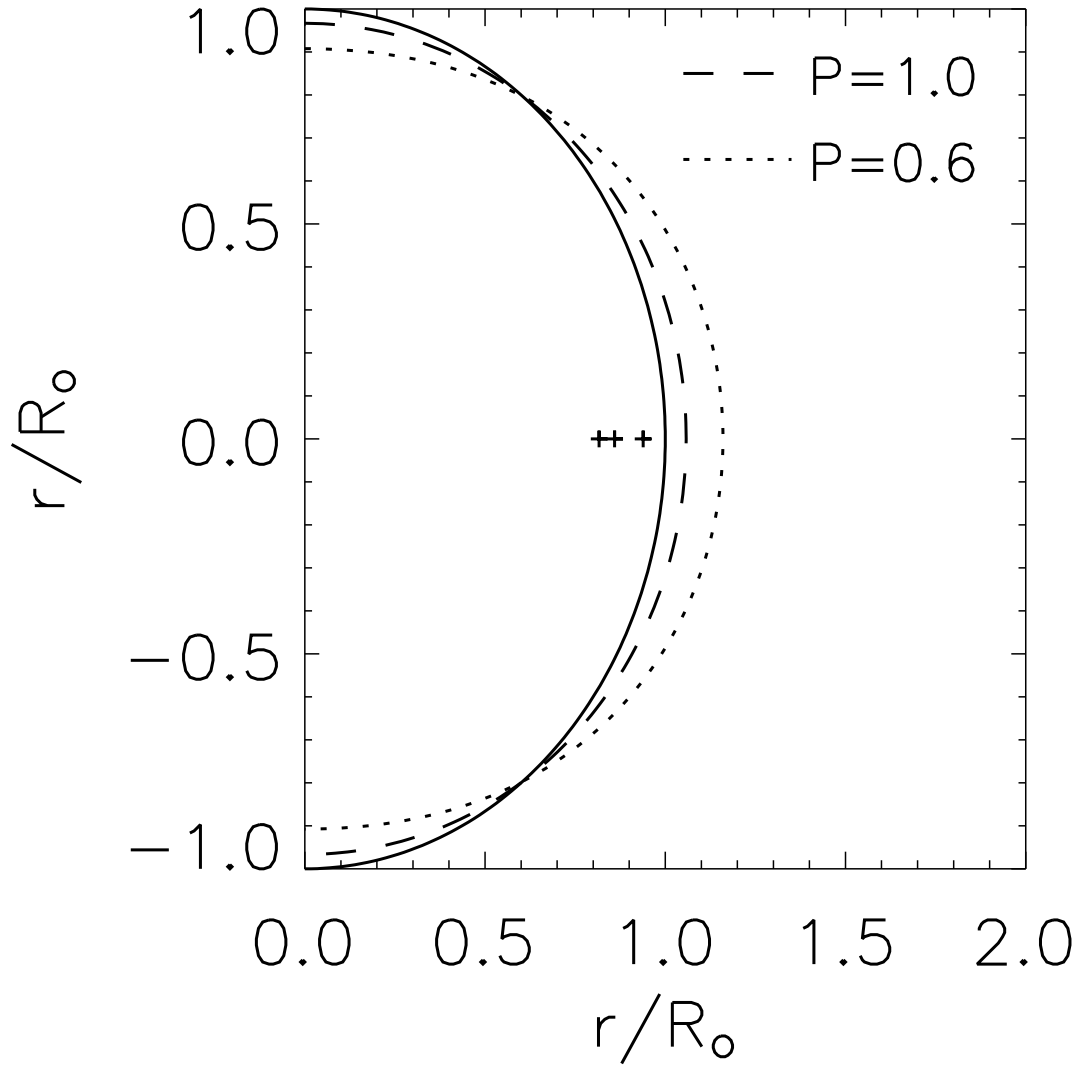}
\caption{The effect of
rotation on the surface shape and radius of gyration (plus sign) of
an n=1.5 polytrope with a central density of 0.8 gm cm$^{-3}$, chosen
to model an ONC star. Two rotation rates are compared with a
non-rotating star. A 0.6 d rotation period corresponds to the maximum
possible rotation rate \citep{j64}. }
\label{rotating_polytrope}
\end{figure}

As an illustration of the effect of rotational distortion we show, in
Fig. \ref{rotating_polytrope}, the surface shapes predicted by this
theory for two rotation periods (1 and 0.6 d) of a PMS star. We have
adopted a value of $\rho_c = 0.8$ in cgs units, which is
representative of an ONC star according to the models. The smaller
period value is very close to the maximum rotation rate (P=0.602 d)
allowed for such an object \citep{j64}. It is also quite close to the
maximum observed rotation rate (P=0.66 d) for ONC stars in the mass
range considered here. For our purposes, it is the effect of the
distortion on the calculation of the radius of gyration which is
relevant. Assuming that the surface shell is thin compared to the
radius of the star and of uniform density one can easily integrate
over the surface shape to calculate a value of k. For a perfect
sphere, k = $\sqrt {2/3}$. In general,
$$k^2 = {{{4 \over 3} a^4 + {16 \over 15} a^3 b  + {8 \over 7} a^2
b^2  + {16 \over 105} a b^3 + {52 \over 1155} b^4}  \over {2 a^2 + {2
\over5} b^2} } $$
Values of k for the three shape solutions are shown as plus signs on
Fig. \ref{rotating_polytrope}.
The shapes for ZAMS stars (n=3) are essentially the same but because
of higher central densities, significant distortion occurs only at
rotation periods significantly shorter than 1 d.

\begin{figure}
\plotone{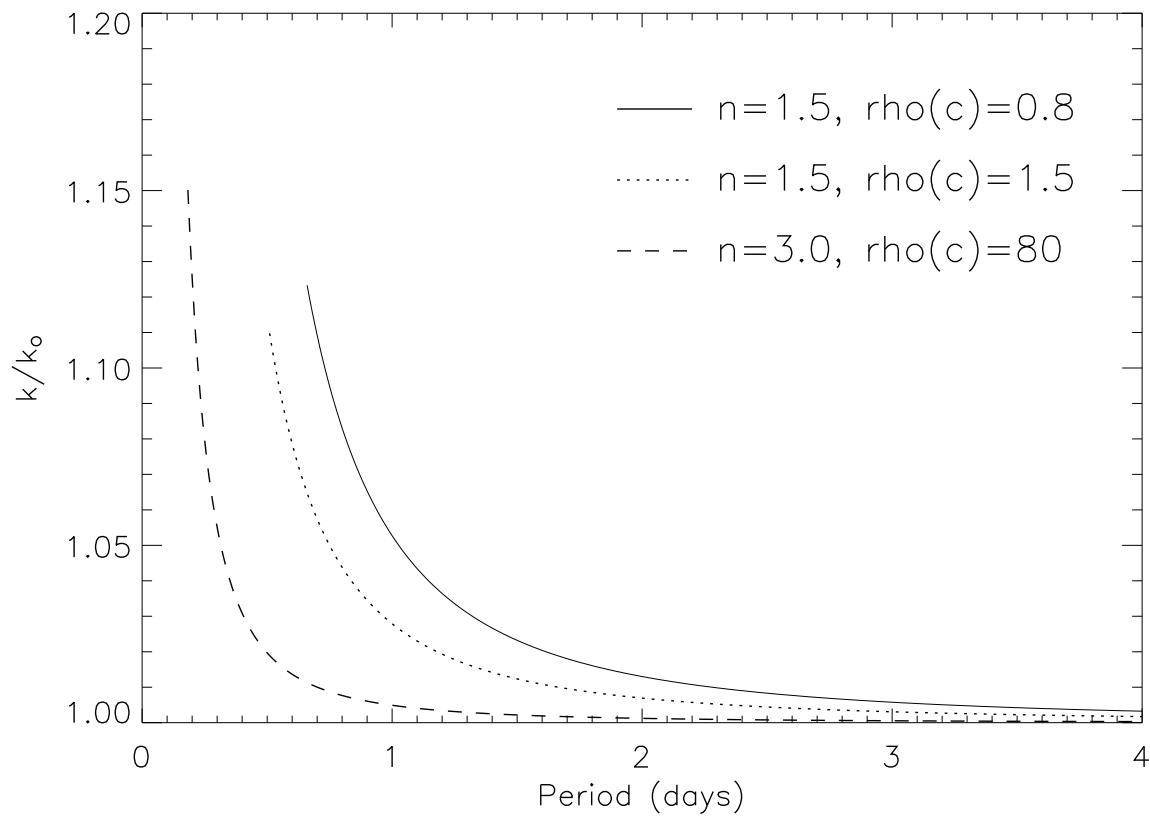}
\caption{The variation of
the normalized radius of gyration (kR$_o$/k$_o$R$_o$) as a function
of rotation period for the three cases indicated. These represent
ONC, NGC 2264 and recently arrived MS stars, respectively. In each
case, the sequences are terminated at the shortest observed period,
which is quite close to the expected shortest period based on the
theory of \citet{c35} and \citep{j64}. 
}
\label{gyration_radius}
\end{figure}

Clearly rotational distortion is not important for most of our sample
but it is important for the most rapidly rotating stars. As noted,
the relevant quantity is k, which enters as the second power in the
calculation of j. To assess and correct for the rotational flattening
we have calculated k by the above formulation. Stellar models suggest
$\rho_c = 0.8$ for ONC stars $\rho_c = 1.5$ for NGC 2264 stars and
$\rho_c = 80$ for ZAMS stars, which we adopt. Fig.
\ref{gyration_radius} shows how k varies with rotation period for
models representing the ONC, NGC 2264 and the ZAMS stars
respectively. In each case, we have terminated the calculations at
the location represented by the most rapidly rotating star actually
observed in each of these clusters (within the mass range considered
here). It is, perhaps, worth noting that in each case the shortest
period star observed is well matched to the shortest period expected
based on \citet{j64} calculations. Stars with shorter periods would
have surface gravities at their equators which were less than zero.
Clearly, neglecting rotational flattening in calculating j for the
most rapidly rotating stars in our sample would lead to errors as
large as 30\%. It is also clear, however, that for most stars the
correction for flattening will be trivially small. 

\section{Results}

Employing the principles outlined in Section II we now derive and
discuss the distributions of P, R and j for solar-like stars at three
different characteristic ages: 1, 2 and 50 My. This is followed by a
discussion, in Section IV, of the evolution of j.

\subsection{Rotation Period Distributions}

\begin{figure}
\plotone{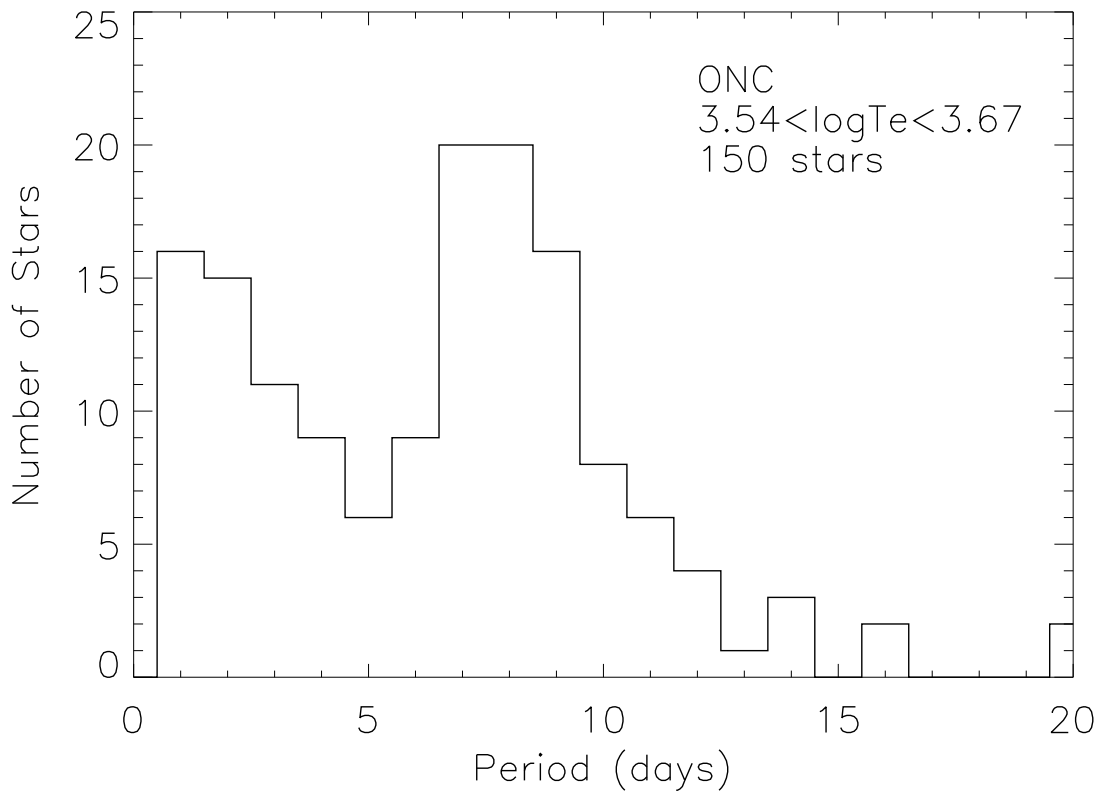}
\caption{Rotation periods
for stars in the ONC with spectral types (K4-M2) appropriate to the
log T$_{eff}$ range of 3.54 - 3.67. Periods are based on the work of
\citet{smm99} and \citet{hbm02} as summarized in the latter paper.
Effective temperatures are assessed by spectral type, as reported by
\citet{h97}, and employ the calibration of \citet{ck79}.  One star,
with a period of 35 days, lies outside the boundaries of this
figure.}
\label{per_hist_onc}
\end{figure}

It is clear from the discussion above that the form of the
j-distribution for young clusters is primarily determined by the
distribution of rotation periods because radii vary little in
comparison to periods. It is, therefore, instructive to look first at
the rotation period distributions of the clusters. Fig.
\ref{per_hist_onc} shows the rotation period distribution for the 150
stars within the selected T$_{eff}$ range in the ONC, which
corresponds to mass between 0.4 and 1.2 M$_\odot$. It has the
familiar bimodal character first reported by \citet{ah92}, with peaks
near 2 days and 8 days. We emphasize that this figure contains every
published rotation period for members of the ONC, regardless of
source, but does not include periods for stars in Orion outside of
the ONC. The reason, again, is that in our view such stars are not
likely to be of the same age and, therefore, radius, as the ONC
members. ``Orion" stars which are members of the flanking fields or
the greater Orion association are likely to be a heterogeneous mix of
stars of different age, mostly older than the ONC, in our view.
Therefore, one would not expect them to exhibit as much structure
(i.e. bimodality) in their period distributions and, in general, one
would expect more rapid rotators. This is, in fact, exactly what is
reported by \citet{r01} for the ``flanking fields" and by
\cite{chs01} for the greater Orion association. 

\begin{figure}
\plotone{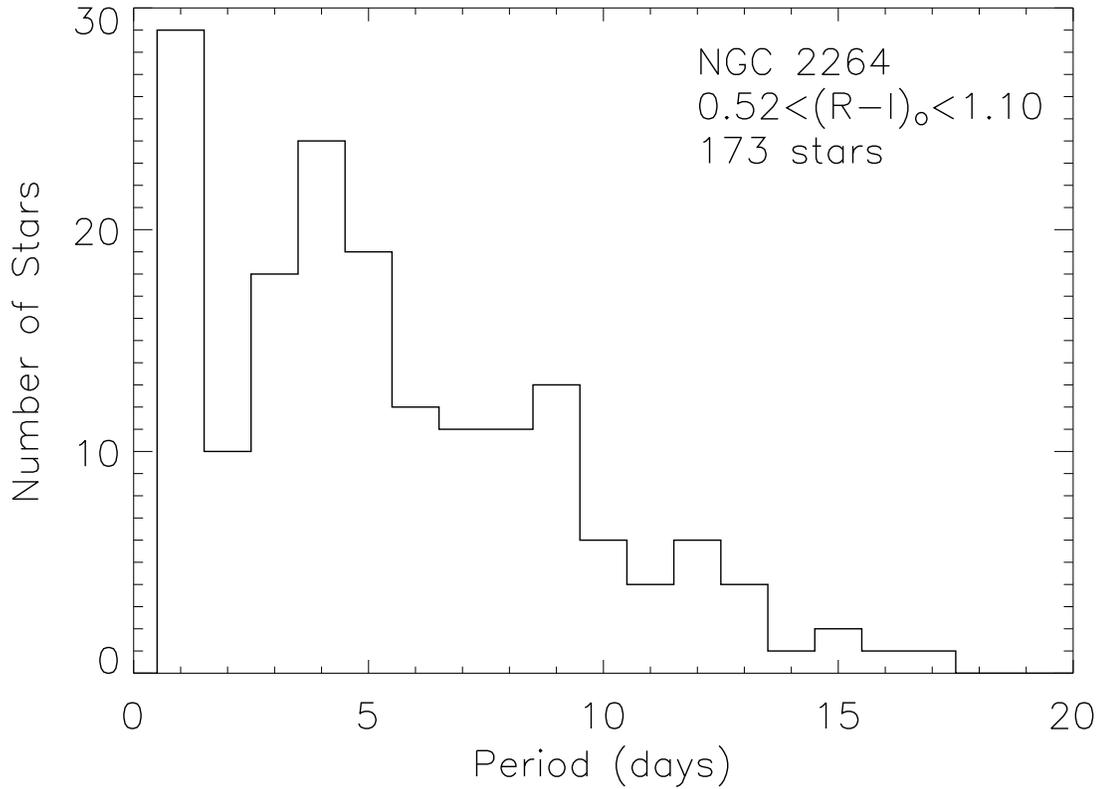}
\caption{Rotation periods
for stars in NGC 2264 with R-I values appropriate to the log
T$_{eff}$ range of 3.54 - 3.67. The periods come from \citet{lbm04}
for 142 stars and \citet{mrs04} for 31 additional stars which were
not detected as periodic by \citet{lbm04}. Only quality 1 stars from
M04 were used. This distribution differs from the one shown in Fig.
\ref{per_hist_onc} for Orion at the 99.7\% confidence limit according
to a K-S test.}
\label{per_hist_2264}
\end{figure}

The rotation period distribution for the 173 stars in NGC 2264 which
lie within the specified color range appropriate to a spectral class
range of K4 - M2 is shown in Fig. \ref{per_hist_2264}.	The 142 stars
with periods detected by \citet{lmb04} and appropriate values of
color were supplemented by 31 stars of quality 1 from \citet{mrs04}.
Reasons for not using quality 2 stars from \citet{mrs04} are given by
 \citet{lmb04}. A double-sided Kolmogorov-Smirnoff (hereafter, K-S)
test shows that there is no significant difference between the
distribution shown in Fig. \ref{per_hist_2264} and the period
distribution for stars in the same color range chosen only from the
sample of  \citet{mrs04}. It is also quite clear that the period
distributions of the ONC and NGC 2264 stars in this mass range are
not drawn from the same parent population. A K-S test indicates that
they are different at the 99.7\% confidence limit. While this
contradicts the statement in \citet{mrs04} that there is no
significant difference between ``Orion" and NGC 2264, it should be
kept in mind that by ``Orion", those authors are generally not
referring to the ONC but to the greater Orion association. In fact,
as Fig. 7b of \citet{mrs04} shows, their period distribution in NGC
2264 does differ from that in the ONC at the 99\% confidence limit
when a K-S test is applied. Other features of this distribution have
been discussed by  \citet{lmb04} and include its bimodality, with
peaks near 1 days and 4 days and the extended tail of slowly rotating
stars.

\begin{figure}
\plotone{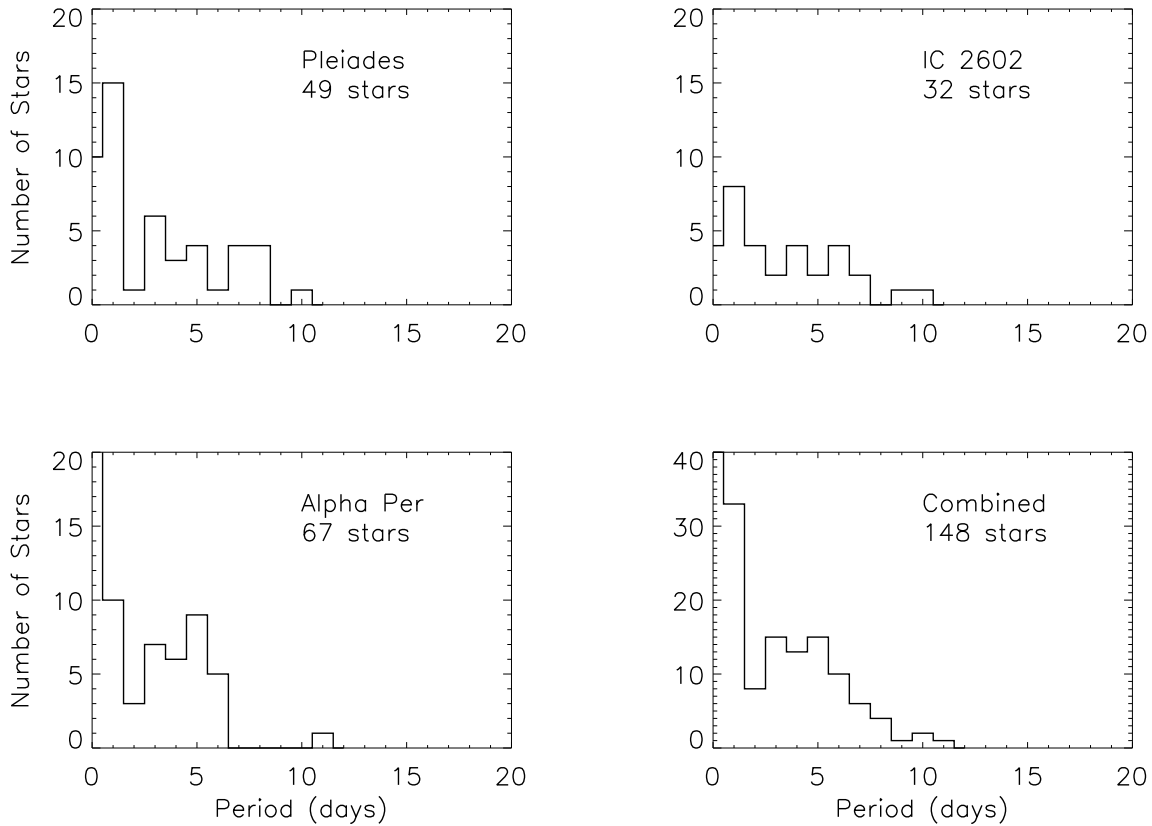}
\caption{Rotation periods for the three clusters containing MS stars
discussed here, from the data in the Prosser catalogue. It is clear
that $\alpha$ Per has more rapid rotators than the others; a K-S test
shows that it differs at the 98\% confidence limit from the
distribution for the Pleiades and at the 99\% confidence limit from
IC 2602, which themselves are not significantly different according
to the same test. We show the combined distribution only for
illustrative purposes since a proper combination requires some
correction for wind losses. One must also keep in mind that selection
effects may be influencing these distributions; in particular, the
Pleiades sample is likely to be biased towards rapid rotators as
discussed in the text. Note the disappearance of the very slow rotators seen in the ONC and NGC 2264. }
\label{per_hist_ms}
\end{figure}

Rotation periods for the three MS clusters are shown in Fig.
\ref{per_hist_ms}. Combined, there are 148 stars, enough to
reasonably define the distribution in a statistical sense. However,
it is not entirely valid to simply combine these three clusters,
because they do not have the same period distributions, as is evident
from the figures and confirmed by a K-S test. Clearly, $\alpha$ Per
has a higher proportion of rapid rotators than do the other two
clusters. From a strictly empirical point of view, this could be
caused by an age difference between the clusters and a general
slow-down of rotation with age expected from wind losses, by
mass-dependent rotation properties and a difference in the mass
distributions between the clusters, or by some other selection
effect. It is easier to explore these issues in the j plane than in
the P plane, so we postpone the task of combining the data until
after a discussion of radii. The combined period distribution shown
in the bottom right panel of Fig. \ref{per_hist_ms} is not strictly
valid given the real differences between the clusters. However,
because these differences are relatively small and a main feature of
the combined plot, namely, its evident bimodal nature is worth
noting, we show the distribution as a didactic exercise. 

To summarize, even without correcting for the effects of radius, a
few things are clear about the rotation distributions of PMS and
recently arrived MS stars of solar-like mass. First, there are
indications on Fig. \ref{per_hist_ms} that the rotational bimodality
observed for the ONC and NGC 2264 continues into the early MS phase;
this will become more evident when the j-distributions are discussed
below. Second, the period distributions are significantly different
from one another at each age step, and third, the trend is for most
stars to spin faster as they age, exactly as one would expect if
angular momentum conservation were involved, at least to some degree.
To assess things more physically and quantitatively we need to
examine j rather than P. This, in turn, requires that we take account
of the stellar radii, a task to which we now turn.
    
\subsection{Stellar Radii}

\begin{figure}
\plotone{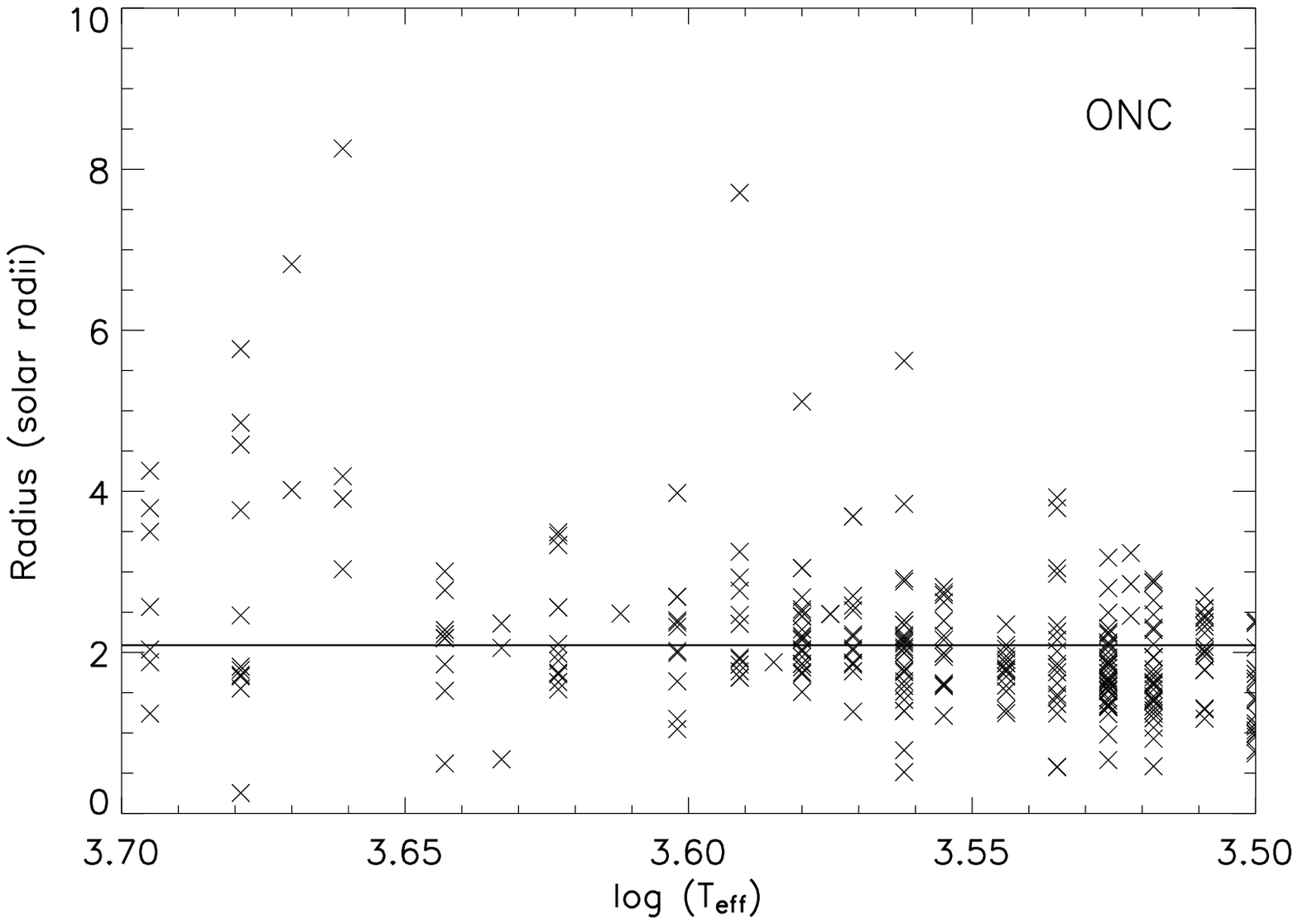}
\caption{Radii (in solar
units) of ONC stars with known rotation periods as a function of
effective temperature. Radii and temperatures are from \citet{h97}.
The median value of the 150 stars with log T$_{eff}$ between 3.54 and
3.67 is R = 2.09 and is shown on the plot and adopted in our
analysis.}
\label{logte_radius_onc}
\end{figure}

Fig. \ref{logte_radius_onc} shows the distribution of radii as a
function of log T$_{eff}$ for ONC stars in our periodic sample. As
expected, there is a wide scatter but no clearly evident trend with
temperature. All data were taken directly from \citet{h97}. Taken at
face value, the wide range of radii would indicate stellar ages that
range from about 0.1 to 10 My \citep{ps99}. As noted previously, our
position is that this scatter is dominated by errors and that the
actual age (and, therefore, radius) spread in the ONC is probably
quite small. Since there is no clear trend of R with T$_{eff}$
visible in the data, we will adopt the median radius of R=2.09
R$_\odot$ for all stars. This is a more robust value than the mean
(2.3 $\pm$ 0.1) because of the outliers at large radius. We show
below that adopting a single value of R=2.09, as opposed to
individual radii, has no effect on the calculated j-distribution
other than to tighten it. Since rotation periods are very accurately
determined and have a large range, while radii are evidently poorly
determined but expected to have a very small range, we argue that
this is the most appropriate procedure if the intention is to obtain
the best estimate of the j-distribution of a cluster population. 

\begin{figure}
\plotone{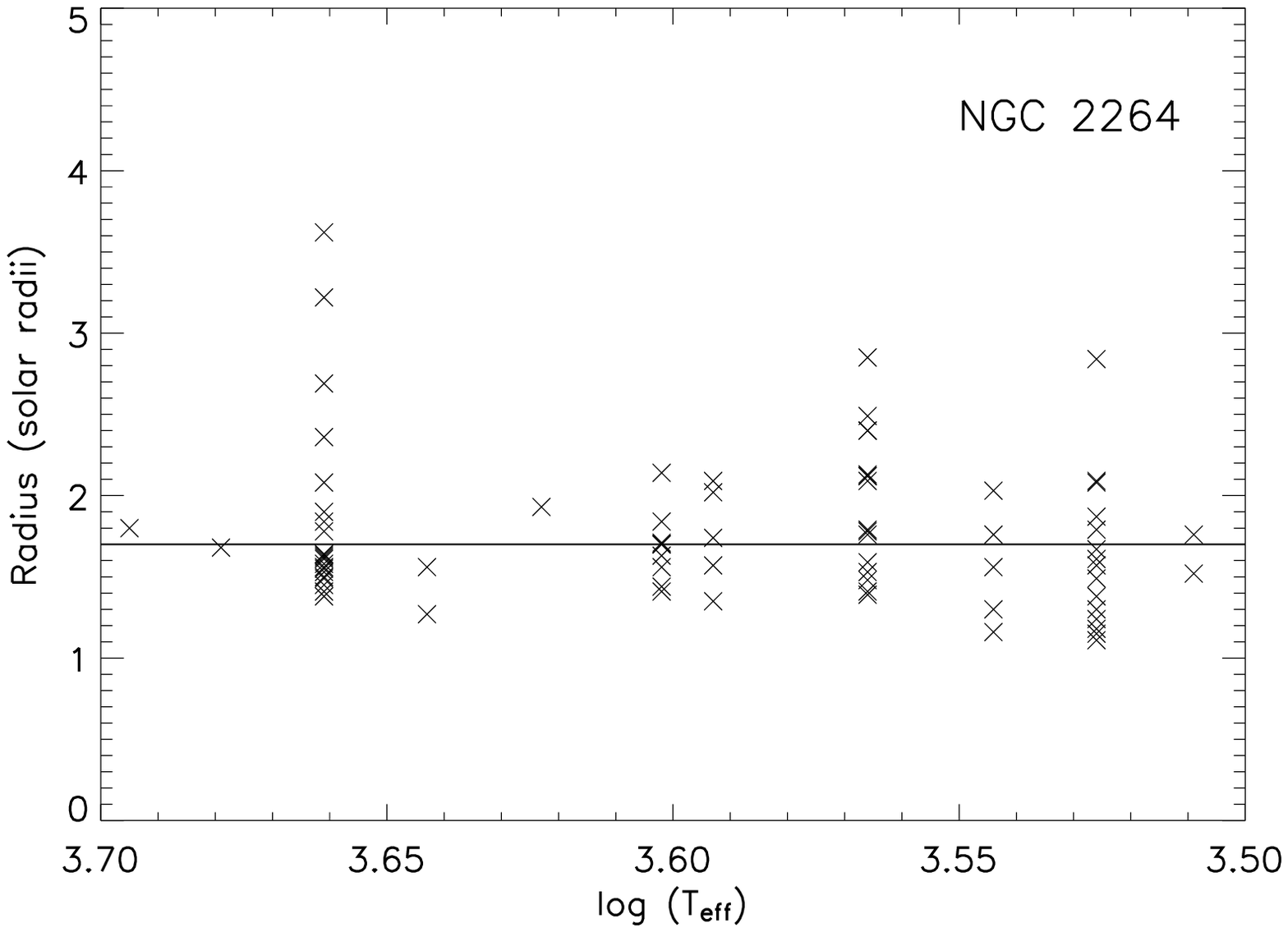}
\caption{Radii (in solar
units) of NGC 2264 stars with known rotation periods from
\citet{lbm04} as a function of effective temperature. Radii and
temperatures are from \citet{rws02}. The median value of the 60 stars
with log T$_{eff}$ between 3.54 and 3.67 is R = 1.70 and is shown on
the plot and adopted in our analysis.}
\label{logte_radius_2264}
\end{figure}

In Fig. \ref{logte_radius_2264} we show the radii of NGC 2264 stars
in our sample. Only 60 of the 173 periodic stars have radius
estimates because spectral types are not available for the rest.
Radii are based on the data and procedures of \citet{rws02} which are
identical to what \citet{h97} has employed in the ONC. There should
be no systematic errors introduced by this procedure, therefore.
Again, we adopt the median radius of 1.70 R$_\odot$ rather than the
mean (1.81 $\pm$ 0.03) to minimize the effect of outliers at large
radius. It is clear that, within the adopted range, stars of the same
spectral class in the ONC are generally larger than those in NGC 2264
by about 25\%. During the Hayashi phase R depends on age (t) as R
$\alpha$ t$^{-1/3}$, so this implies that the ONC is about one-half
the age of NGC 2264. A fiducial age of 1 My for the ONC implies an
age of 2 My for NGC 2264. The fact that NGC 2264 is somewhat older
than the ONC, based on the observation that stars of the same
effective temperature are somewhat less luminous, is now well
documented in the literature \citep{mrs04, lbm04}. As in the ONC, we
find no evidence for a dependence of R on log T$_{eff}$ over the
limited range of interest in this study. 

\begin{figure}
\plotone{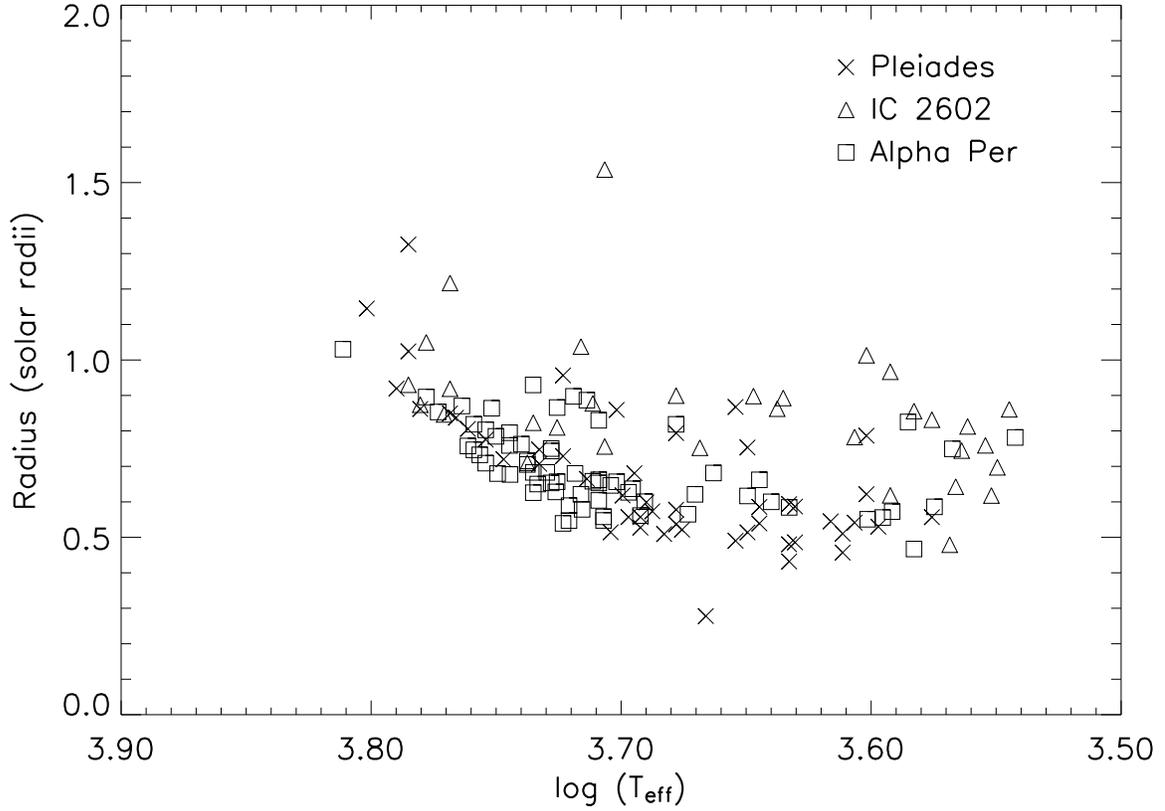}
\caption{Radii (in
solar units) of the Pleiades (X's), IC 2602 (triangles) and $\alpha$
Per (squares). The effect of unresolved binaries on the determination
of radii is apparent; a clear binary sequence exists above the single
star MS. There is one discrepant point in the Pleiades (HII 1280, a
K7 star with one of the shortest rotation periods measured in the
cluster --- 7.25 hours) and one obviously discrepant star in IC 2602
(B1 34, which has one of the longest rotation periods in the
cluster).}
\label{logte_radius_ms}
\end{figure}

Radii for stars in the MS clusters can be determined much more
accurately than for PMS stars, as discussed above. Hence, we use
individual values of radius in calculating j for these stars. The
distribution of radii for each cluster is shown in Fig.
\ref{logte_radius_ms}. In general, the stars describe a very tight
(main) sequence with a parallel binary sequence above it. There are
only two widely discrepant points. One, in the Pleiades, is the star
HII 1280, a K7 star with one of the shortest rotation periods
measured (7.25 hours). Its radius is well below the MS because its
measured color is too blue for its brightness. The cause of this
discrepancy is unknown but could be related to its extreme rotation.
The one discrepant star in IC 2602 is B1 34, with a radius clearly
too large for its effective temperature. It is interesting that it
has one of the longest rotation periods in the cluster (P = 6.7
days). The presence of two outliers has no effect on the analysis of
this paper so we simply include them as interesting anomalies. As
discussed above, we deliberately neglect the binary sequence (by treating them as single stars) in
discussing the evolution of j because no correction for binaries can
be made for the PMS clusters. We show in what follows that the effect
of the binaries on the j distribution of the MS clusters is small. 

A visual summary of this section is given in the right panel of Fig.
\ref{hr_diagram}. The solid lines indicate the median radii for the
PMS clusters and individual radii are plotted for the MS cluster
members. It is interesting to note that although the age difference
between the ONC and NGC 2264 is quite small compared to the age
difference between either cluster and the MS stars, the radius
difference is relatively more substantial. In other words, the rapid
evolution of radius during the early PMS phase is quite clear on this
figure and an important factor to keep in mind when interpreting
rotation data of PMS stars, especially if a population of
heterogeneous age is considered. With P and R distributions now in
hand it is possible to move on to the physically more relevant
quantity, j.

\subsection{j distributions}

\begin{figure}
\plotone{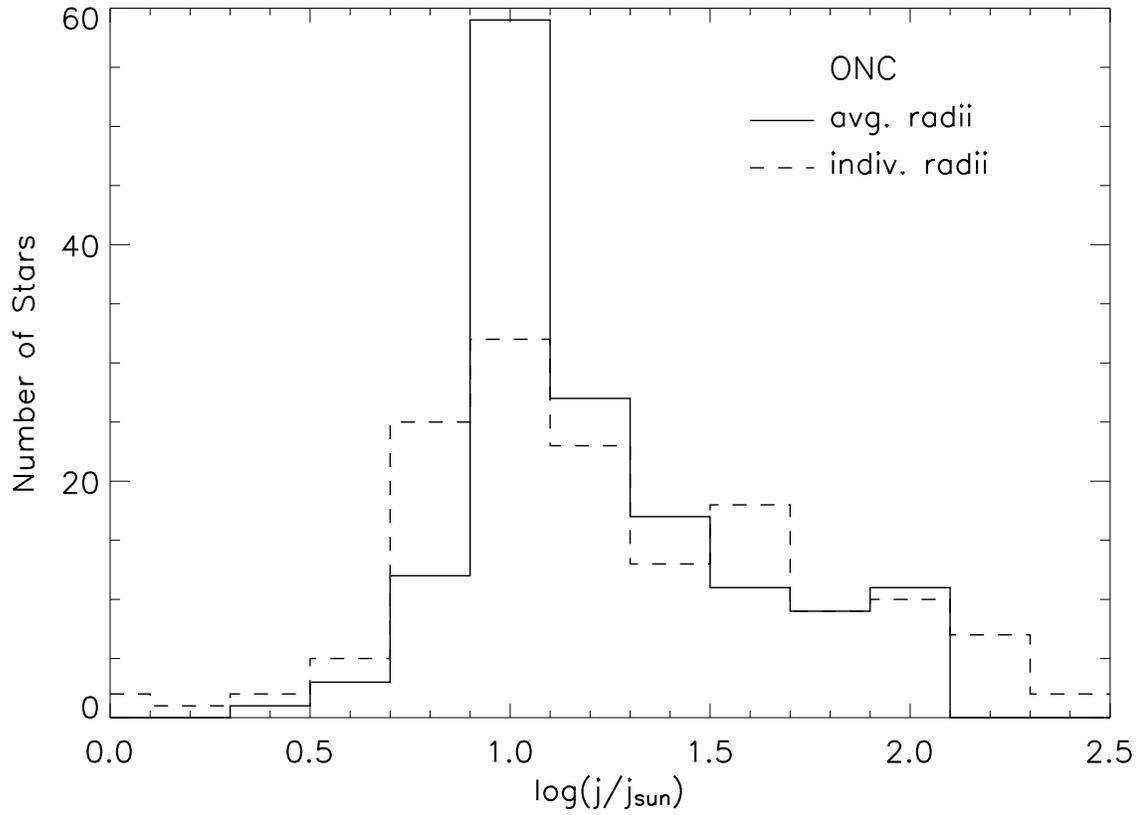}
\caption{The frequency 
distribution of specific angular momentum (j) in solar units for
stars in the ONC with log T$_{eff}$ between 3.54 and 3.67. The
distributions are calculated using a mean radius of 2.09 solar radii
(solid line), and using individual radii (dotted line). As expected,
the distribution using a mean radius is tighter than that using
individual radii and there is no systematic shift.}
\label{j_hist_onc}
\end{figure}

The calculation of j follows directly from P and R. For convenience
we normalize the results to j for the Sun (j$_\odot$), which is based
on an adopted solar radius of 6.96 x 10$^{10}$ cm and a mean surface
rotation period of 25 days. For a spherical shell, k has the value
(2/3)$^{0.5}$ so j$_\odot$ = 9.4 x 10$^{15}$ cm$^2$ s$^{-1}$. To
begin, we computed j for the ONC stars in two ways, using the
individual radii and using the median radius of 2.09 R$_\odot$. A
comparison of the resulting j distributions is shown in Fig.
\ref{j_hist_onc}. As expected there is no systematic difference
between these
but the distribution based on the median radius is tighter. As argued
above, we believe the large scatter in the  radii in the PMS clusters
is primarily due to errors in their determination, not to real
variation, so we are not surprised that the distribution based on
individual radii is broader. It simply reflects an additional source
of scatter, in our view. In what follows we use only the distribution
based on the mean radii, both for the ONC and for NGC 2264. 

\begin{figure}
\plotone{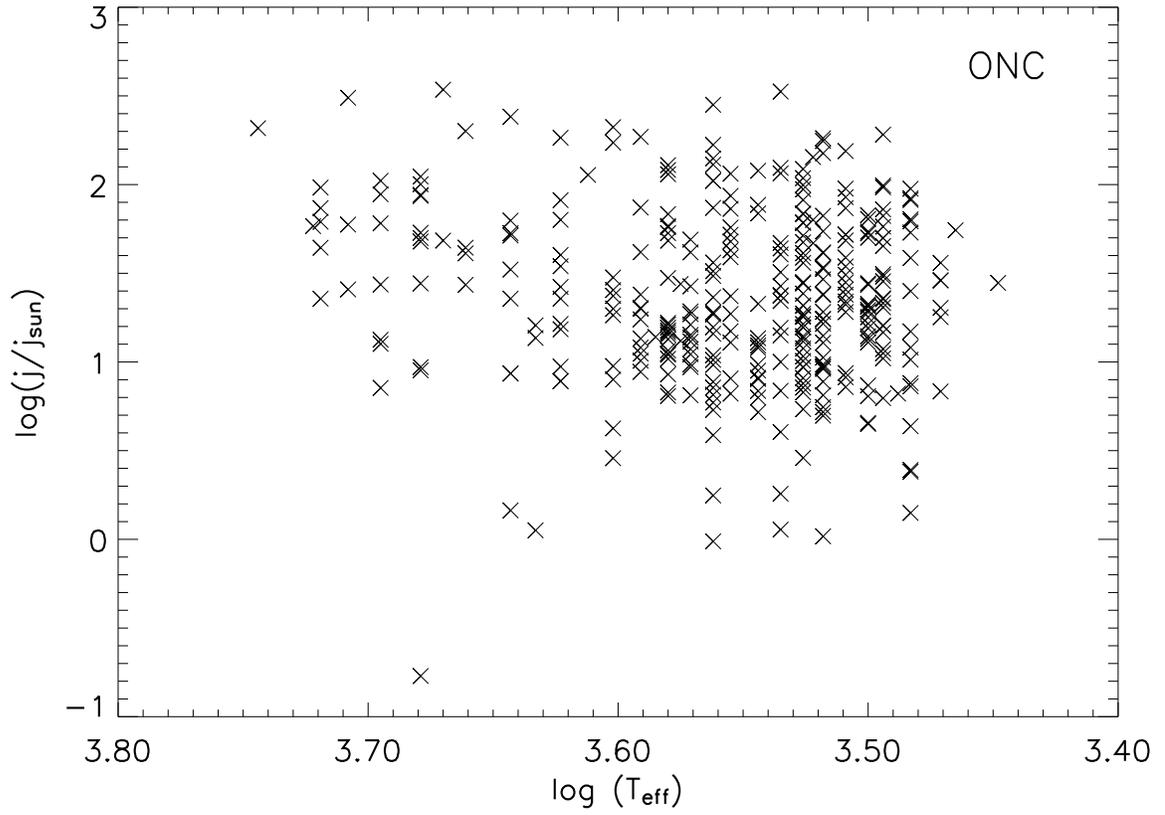}
\caption{The distribution
of specific angular momentum (j) in solar units for stars in the ONC
versus effective temperature. It is clear that there is a wide
distribution of j at all temperatures (masses) and that little or no
trend of j with mass is apparent.}
\label{j_logte_onc}
\end{figure}

\citet{hbm01} showed that although the P distribution is a function
of mass in the ONC, the j distribution is nearly independent of mass
over the range 0.1 - 1.5 M$_\odot$. The more rapid rotation
characteristic of lower mass stars (outside the mass range considered
here) is compensated for by their smaller radii. For the more limited
mass range considered here it is not surprising, therefore, to find
no evidence for a dependence of j on T$_{eff}$, as illustrated in
Fig. \ref{j_logte_onc}. Similarly, there is no evidence for a
dependence of j on effective temperature in NGC 2264, although only
about 1/3 of the sample has known spectral type.

\begin{figure}
\plotone{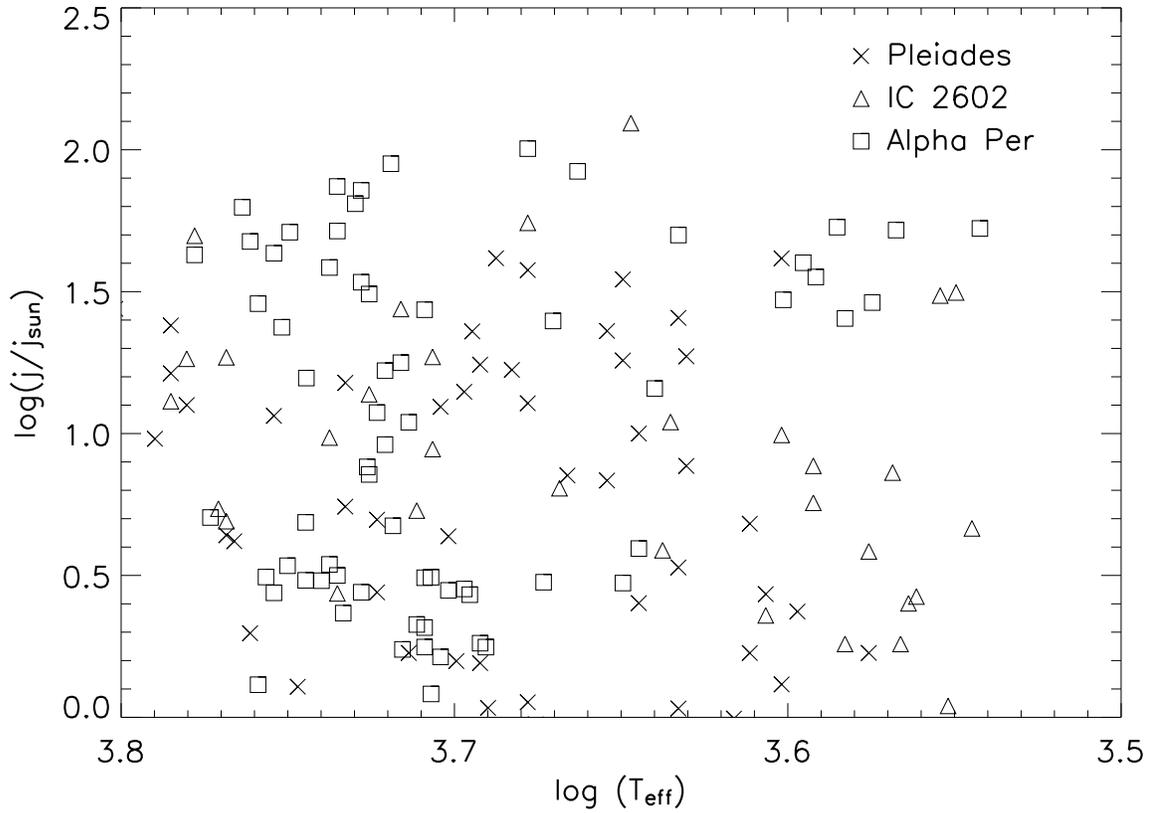}
\caption{The distribution of
specific angular momentum (j) in solar units for stars in the
Pleiades, IC 2602 and $\alpha$ Per clusters versus effective
temperature. It is clear that there is a wide distribution of j at
all temperatures (masses) and that little or no trend of j with mass
is apparent. There are small differences between the clusters which
may or may not be significant, as discussed in the text. In
particular, $\alpha$ Per has a set of 8 low mass stars which are all
rapidly rotating and no slow rotators of comparable mass. It also has
a greater proportion of rapid rotators at all masses than the
Pleiades.}
\label{j_logte_ms}
\end{figure}

\begin{figure}
\plotone{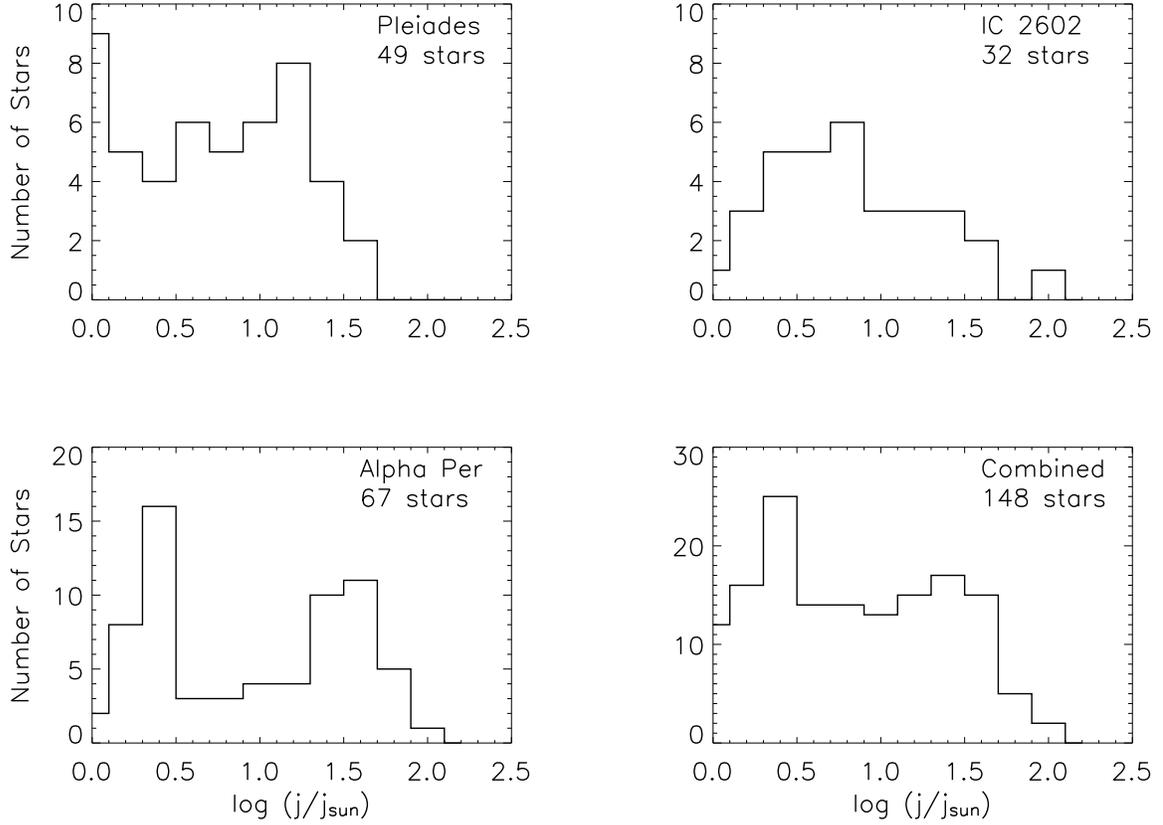}
\caption{The frequency	distribution of specific angular momentum (j)
in solar units for stars in the Pleiades, IC 2602 and $\alpha$ Per. A
K-S test indicates that the parent populations of the $\alpha$ Per
and Pleiades samples do not have the same rotation properties at a
confidence level of 99\%. IC 2602 is intermediate in its properties
and, given the relatively small number of stars, inconsistent with
both other clusters at only the 1-2 $\sigma$ level. It is reasonable
to ascribe the generally slower rotation of stars in the Pleiades to
the effect of stellar winds acting over the 50-70 million years that
separates the clusters in age but selection effects, as discussed in
the text, may also play an important role.}
\label{j_hist_ms}
\end{figure}

Fig. \ref{j_logte_ms} shows the values of j calculated for stars in
the MS clusters, again as a function of T$_{eff}$. Overall it is
clear that there is little or no trend seen here either. However,
there are some differences between the clusters. In particular,
$\alpha$ Per contains a group of eight high j stars at the low mass
end of the distribution. It is evident from the HR diagram (Fig.
\ref{hr_diagram}) that these are also not fully contracted to the MS.
There is no corresponding set of more slowly rotating stars. It is
hard to say whether this is a real, significant difference given the
small number of stars involved. A K-S test does show that, like the
rotation period distribution, the j distribution of $\alpha$ Per is
significantly different from the Pleiades (and IC 2602). This is
shown clearly in Fig. \ref{j_hist_ms} where frequency distributions
for all three clusters are displayed. The Pleiades distribution
differs from  $\alpha$ Per at the 99\% confidence level, containing
more low j stars. IC 2602 is intermediate in its properties (and
contains less stars), differing from each of the other clusters at
only the 1-2 $\sigma$ level. The combined j distribution is shown in
the bottom right hand panel for illustrative purposes only. It is
clearly bimodal, reflecting the bimodal period distributions of both
the Pleiades and $\alpha$ Per.

From a purely empirical view it is not entirely appropriate to
combine the j distributions of the three MS clusters since there is
evidence that they were not drawn from the same parent population. 
One interpretation is that the ages of the three clusters are
sufficiently different that the action of normal stellar winds over
the time interval between them is sufficient to have measurably
slowed the Pleiades stars with respect to the $\alpha$ Per and IC
2602 stars. Another is that there are selection effects that are
biasing the distributions. A third is that the clusters simply did
not begin their lives with the same initial j distributions.
Unfortunately, there is no way of knowing for sure which of these
possible effects is, indeed, important but fortunately the
differences among the MS clusters are small compared to the
differences between the PMS and MS clusters. 

\begin{figure}
\plotone{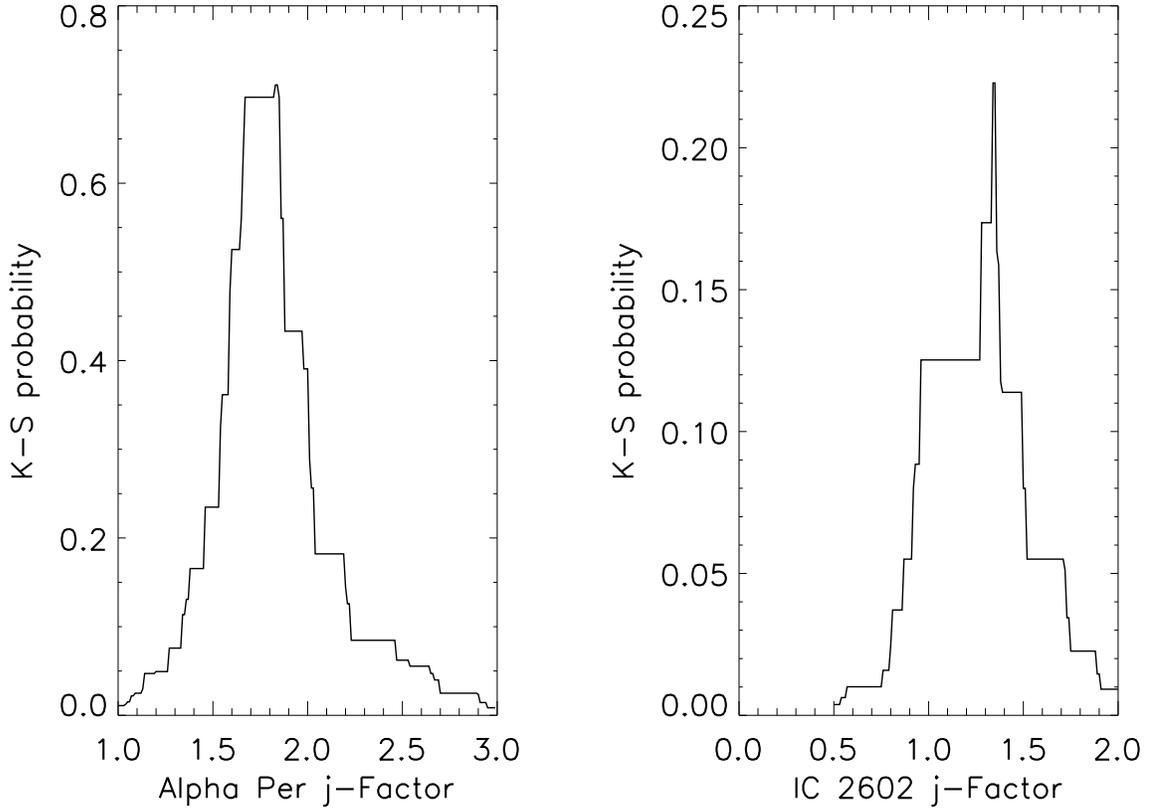}
\caption{The
probability, as calculated by a double-sided Kolmogorov-Smirnoff
test, that the j-distribution of the Pleiades and $\alpha$ Per (left
panel) clusters have the same parent population, as a function of the
factor by which the Pleiades distribution is scaled. In other words,
if j for every star in the Pleiades is multiplied by a factor of
about 1.75 the distribution is statistically indistinguishable from
what is observed for the $\alpha$ Per cluster. The corresponding
factor for agreement between IC 2602 and the Pleiades is 1.35. More
significantly, the range of j-factors that give satisfactory fits by
this simple scaling process at the 2$\sigma$ level is 1.3 - 2.2 for 
$\alpha$ Per and 0.9 - 1.5 for IC 2602. See text for further
discussion of the implications of this figure.}
\label{age_factors}
\end{figure}

We proceed empirically by asking whether there is a simple
transformation of the MS cluster j-distributions that leads to
statistically acceptable agreement among them. The Pleiades
distribution is taken as the fiducial point, and we seek to transform
its j-distribution to each of the others by applying a constant scale
factor (the ``j-factor"). The results are shown in Fig
\ref{age_factors}. It is clear from this exercise that if the
Pleiades stars with known rotation periods all had about 1.75 times
more angular momentum per unit mass than their counterparts in the
$\alpha$ Per cluster, the distributions would be statistically
indistinguishable from one another. The corresponding factor for best
transforming the Pleiades j-distribution to the IC 2602 distribution
is about 1.35. In both cases, we find that the demonstrably older
cluster, the Pleiades, has lower j-values than the younger MS
clusters, in agreement with the common supposition that wind losses
are draining some angular momentum from these young stars on a time
scale of tens of millions of years. 

We can assess the situation a bit more quantitatively by assuming
that a Skumanich-type \citep{s72} wind loss relation  ($\omega$
$\alpha$ t$^{-1/2}$) applies to all stars. In that case, a loss by a
factor of 1.75 in j would imply an aging by a factor of 3, indicating
a current age for $\alpha$ Per of 40 My if the Pleiades is 120 My
old. Similarly, we would compute an age of 65 My for IC 2602 by this
process. These are reasonably consistent with the quoted ages of the
clusters given above although the order by age is not correct, and
there is little doubt that $\alpha$ Per is older than IC 2602. We
attribute this small inconsistency in the rotation properties of
these clusters to the relatively small number of stars with known
rotation properties and to possible selection effects in the data
discussed below. At the 2$\sigma$ level (K-S factor $>$ 0.1) we find
agreement in the j-distributions of $\alpha$ Per and the Pleiades for
j-factors of 1.3-2.2, corresponding to ``Skumanic" ages of 25-70 My
for that cluster. For IC 2602, the corresponding numbers are a
j-factor of 0.9-1.5, and an inferred age of 50-150 My. Since this
simple empirical scaling process has neglected complications such as
saturated winds that are probably of importance for the more rapid
rotators in our sample, it is actually remarkable that we get as good
an agreement as we do with ages inferred by more accurate methods.  

\begin{figure}
\plotone{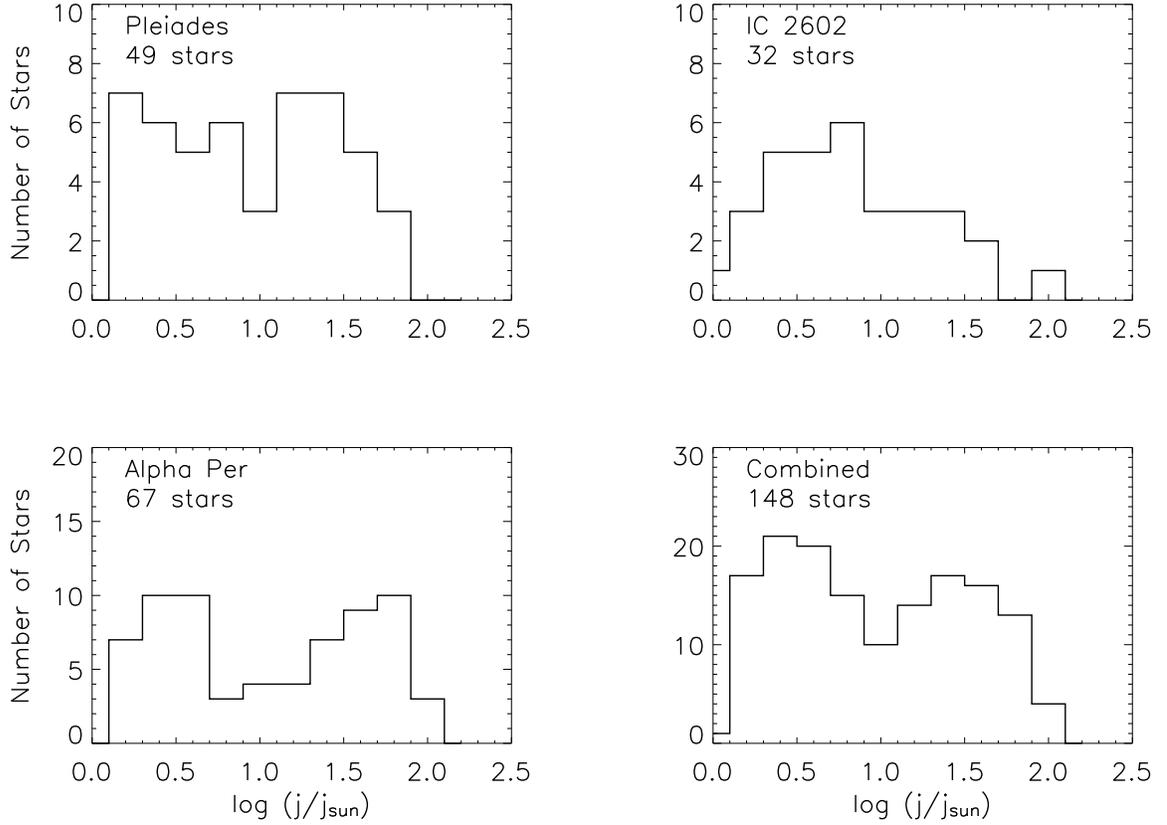}
\caption{The
j-dstributions of the three clusters with MS stars shifted to a
common age of $\sim$50 My (i.e. the age of IC 2602) using the
j-factors based on Fig. \ref{age_factors}. No correction is made for
binary stars because binaries cannot be identified or corrected for
in the clusters containing PMS stars. They have been treated as if they were single stars.}
\label{j_hist_ms_corrected}
\end{figure}

\begin{figure}
\plotone{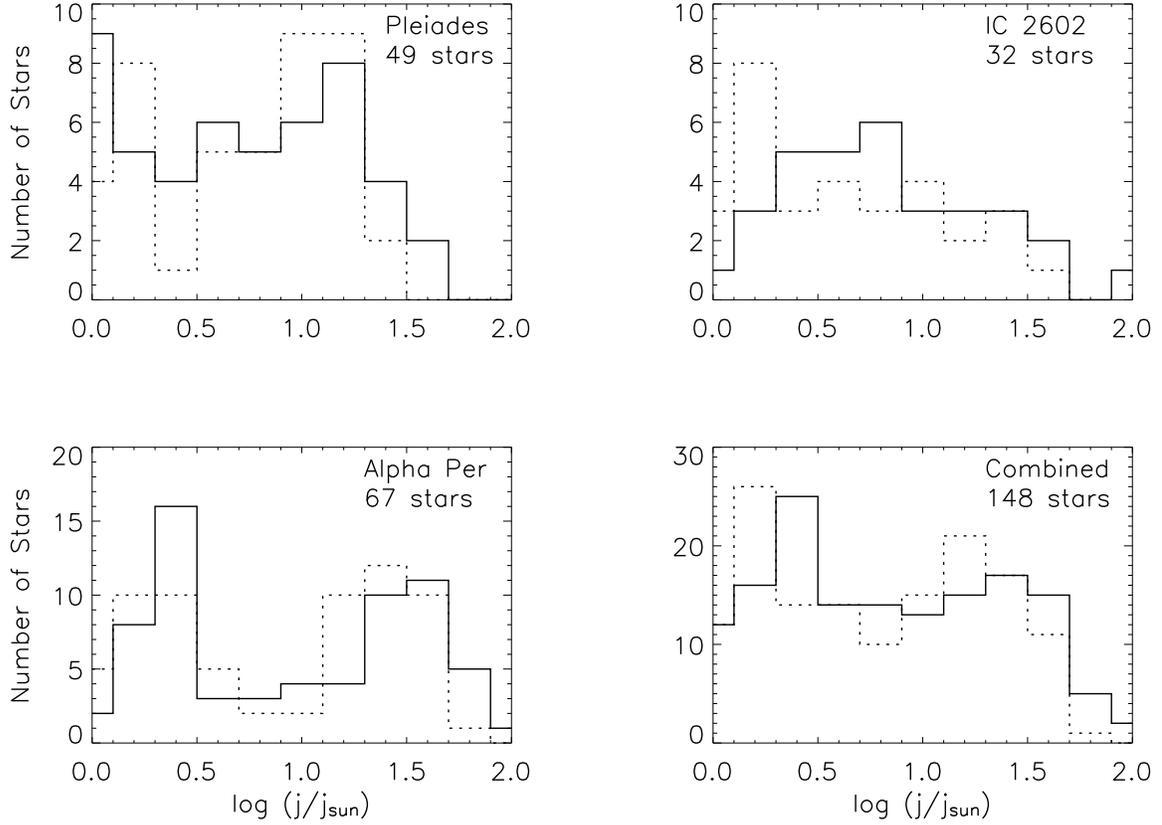}
\caption{The
j-distributions of the three older clusters corrected for the
presence of binary stars. This was done by fitting a line to the
single star sequence of radius versus log T$_{eff}$ and applying that
radius to all stars. The dashed lines show the j-distributions for
the binary-corrected sample, while the solid lines show the observed
j-distributions without a binary correction. As may be seen, the
presence of binaries makes a small difference. We have not attempted
to use this correction in the analysis because there is no way to
correct the PMS stars for this effect. We assume, therefore, that the
binaries have a roughly equal (small) effect on the j-distributions
at all ages and may safely be ignored.}
\label{j_hist_ms_binary_corrected}
\end{figure}

In Fig. \ref{j_hist_ms_corrected} we show the transformed j
distributions of the MS clusters corrected for angular momentum
(presumably wind) losses and adjusted to a common age, namely the age
of IC 2602 ($\sim$50 My). These j distributions are now sufficiently
similar to have been drawn from the same parent population and can be
combined. The bimodal nature of the combined j distribution continues
to be quite clear. We take this combination of the adjusted rotation
distributions to be representative of stars in this mass range at
about the time they arrive on the MS. Note that we have not corrected
the distribution for the presence of binary stars. For illustrative
purposes only, we show the effect of such a correction in Fig.
\ref{j_hist_ms_binary_corrected}. There would be a minor shift to
lower values of j. The reason for not including this correction when
discussing the evolution of j distributions is that it is impossible
to make it for the PMS clusters. There, any binary sequence is lost
in the observational scatter. If a correction were made to the MS
clusters but not to the PMS clusters we would clearly not be making a
valid comparison so we adopt the procedure of ignoring the
(relatively small) correction in both cases.	

\section{The Evolution of Rotational Distributions of Solar-Like
Stars}

We have formed three distributions of j representative of a cluster
population of solar-like stars at three different times, nominally 1
My (ONC), 2 My (NGC 2264) and 50 My (combined MS clusters). Fig.
\ref{j_hists_uncorrected} compares these distributions in pairs and
then with all three shown for clarity. K-S tests reveal that each
distribution is significantly different from the others at more than
a 99\% significance level (see Table \ref{data}). While the
distributions are shown as fractions of the total for easy comparison
with one another it should be recalled that there are 150-175 stars
in each so they are reasonably well defined. 

A new and, we believe, significant feature of angular momentum
evolution emerges from this comparison. As seen clearly on Fig.
\ref{j_hists_uncorrected}, the high-j sides of the distributions are
rather similar in all three data sets, while the low-j sides evolve
dramatically as the population ages.  In other words, rapidly
rotating PMS stars appear to evolve with very little additional loss
of angular momentum to the MS, while slowly rotating stars in the ONC
must lose substantial additional amounts as they progress to the MS.
Although quite evident on the figure, we can quantify the result by
dividing the sample into a rapidly rotating half and a slowly
rotating half. Applying the K-S test to each half independently
yields the significance values given in Table \ref{data}. The rapid
rotator side of the distribution shows only small, if any,
indications for evolution with time, while the slow rotator side
evolves dramatically. 

In our opinion, this feature of the evolution of j-distributions
provides dramatic new support for the disk-locking theory of angular
momentum evolution, as we now discuss. An overview of the argument is
as follows. According to the disk-locking theory, the slower rotators
in the ONC should be those still interacting with their disks, while
the rapid rotators should have lost most or all such interaction at
an earlier stage. Assuming that once the influence of a disk on a
star's rotation has waned, it does not tend to reappear, one would
expect rapid rotators at the ONC age to show only small angular
momentum losses as they age further. The high-j side of the
distribution should not evolve much with time, therefore, precisely
as is observed. If large angular momentum losses are to occur in any
stars, it should be the slow rotators, since these are the ones that
still have disks. Again, this is precisely what Fig.
\ref{j_hists_uncorrected} shows. 

A second aspect of rotational evolution revealed on Fig.
\ref{j_hists_uncorrected} that provides additional new support for
the disk-locking theory is the clear difference in j-distributions
seen between ONC and NGC 2264 ages. By its nature, the disk-locking
theory predicts the most rapid evolution of j (recall that j is the
angular momentum per unit mass at the {\it surface} of the star) will
occur during the most rapid contraction phases. The amount of j loss
should scale with radius of the star, not time elapsed. Since
evolution of radius is most rapid during the initial stages of PMS
evolution (see Fig. \ref{hr_diagram}) the disk-locking theory would
predict a detectable evolution of the j-distributions even on the
very short time scale ($\sim$1 My) represented by the difference in
ages between the ONC and NGC 2264. Other theories of angular momentum
loss (e.g. by winds) would predict an evolution that would be more
steady with time and would not lead to detectable differences among
PMS clusters with such similar ages. We now explore these arguments
in more detail, considering first the evolution of the rapid
rotators.  
 
\subsection{Rapid Rotators: Near Conservation of Angular Momentum}

Conservation of angular momentum on Fig. \ref{j_hists_uncorrected}
would be indicated, of course, by no change in the distributions with
time. The good agreement evident among all distributions on the rapid
side, therefore, means that rapid rotators are evolving in a manner
essentially indistinguishable from conservation of angular momentum:
wind losses, or other angular momentum losses are small or
negligible. This means that there is no need to posit any additional
physics other than PMS contraction and conservation of angular
momentum to account for the ``ultrafast rotators" in young clusters,
i.e. the stars populating the high j side of the distribution in the
MS clusters. To quantify this result, we note from 
Table \ref{data} the K-S probability that the large j sides were
drawn from the same parent populations. These probabilities are:
0.38, 0.24 and 0.14 for the ONC-MS, ONC-NGC 2264 and NGC 2264-MS
pairs, respectively. In other words, at the 1-2 $\sigma$ level or
better, there is no disagreement among the distributions on the
rapidly rotating side. Consequently, there is no significant
empirical evidence for any angular momentum loss between the PMS and
MS phases of rapidly rotating stars. 

Of course, some angular momentum loss due to winds would be expected
and can be accommodated by the data. To quantify this we have shifted
the ONC and NGC 2264 j distributions by a series of different factors
and calculated the K-S probability of their fits with the combined MS
distribution. We find that a marginally better fit can be obtained
between the ONC and the MS clusters if the ONC j distribution is
scaled by a factor of 0.95, corresponding to wind losses of 5\%. NGC
2264 does not improve its fit with the MS clusters if scaled by any
factor and at 0.9 the indication is that there is a 98\% chance that
it was drawn from a different parent population than the MS clusters.
Our empirical evidence, therefore, is that the PMS stars lose less
than 10\% of their angular momentum to stellar winds during the first
50 My of their evolution.  

\subsection{Slow Rotators: A Severe Angular Momentum Drain}

Since the overall j distributions are significantly different from
one another at each of the three epochs, while the rapidly rotating
sides of the distributions are not significantly different, it is
clear from the K-S analysis (or just from the appearance of the
distributions on Figs. \ref{j_hists_uncorrected}) that it is the
slowly rotating stars which are evolving in j so dramatically.
Clearly, slow rotators are losing substantial amounts of angular
momentum as they age. To quantify the significance level of the
effect we compare the low j halves of the distributions using the K-S
test (see Table \ref{data}). Even in  the case of the comparison
between NGC 2264 and the ONC, where the age difference is only
$\sim$1 My, there is a highly significant difference in their j
distributions on the low j side. A K-S test indicates that there is
only a 9 x 10$^{-7}$ chance that the distributions have the same
parent populations. The K-S probability is less than 10$^{-16}$ when
the ONC or NGC 2264 is compared to the MS clusters. It is clear from
Fig.  \ref{j_hists_uncorrected}  that slowly rotating stars lose
substantial amounts of angular momentum during their contraction to
the MS. 

It may be seen that no overall scaling of the distributions can
transform one into another. The reason, of course, is that the j
distributions evolve with time not only by shifting their medians
towards lower j but also by broadening dramatically. There is simply
no way to understand the evolution of these distributions with time
without considering the slow rotators separately from the rapid
rotators. This, of course, is what one would expect from a
disk-locking theory of angular momentum evolution. Slow rotators
should be the ones still locked to their disks and, therefore, the
ones which should continue to lose additional amounts of angular
momentum. This is precisely what our data suggest is happening.
Lacking a quantitative theory of disk-locking it is hard to make a
more compelling comparison of the data and theory. However, it is
possible to estimate empirically what sort of evolution is required
under the disk-locking paradigm to account for the observations. That
is done in the next section.

\subsection{A Disk-Locking Model for the Data}

A clear indication from this study is that stars with j $>$ 10
j$_\odot$ in the ONC must evolve with very little angular momentum
loss during the next 50 My if the j distribution is to transform into
that seen for young MS clusters. At the same time, a significant
fraction of stars with j $\le$ 10 j$_\odot$ must lose substantial
amounts of angular momentum, typically a factor of 3, in 50 My to
populate the low j side of the distribution exhibited by the MS
clusters. This angular momentum loss must, furthermore, be initially
rapid to account for the significant evolution towards lower j
already seen in NGC 2264. Clearly, these aspects of the evolution
point to some kind of locking as the physical mechanism, and the correlations suggest disk-locking. Lacking a predictive theory of disk-locking, it is difficult to go much
further with a quantitative comparison, especially since the locking is likely to be imperfect, as has been discussed by \citet{lmb04} We can, however, make some
quantitative comparisons between the theory and data, a task to which
we now turn.

To begin, we inquire whether there is a simple transformation of the
data between the ONC, NGC 2264 and the MS which can account for the
evolution of the j distribution. From an empirical viewpoint, it is
clear that this transformation must be applied only to the low
angular momentum side, otherwise the reasonably good fit that already
exists on the high j side (Table \ref{data}) would be lost. To
explore the simplest possible transformation that might work we
divided the samples into two sets, a low angular momentum group
comprising a fraction (f) of the whole sample and a high angular
momentum group comprising the rest of the sample (a fraction 1-f).
The low angular momentum group was then multiplied a by a j-factor
(obviously less than 1) and compared them with the MS sample using
the K-S test. This simulates continued loss of angular momentum for
the already low-j stars, as expected in the case of disk-locking.
Exploring (f, j-factor) space in this manner led to the discovery
that there is a fairly narrow range in these parameters which does,
in fact, allow one to match distributions across time in a
statistically acceptable way. Our best fits are listed in Table
\ref{model} and shown in Fig. \ref{models}. The parameters adopted
are given on the figures. For the lower right panel we combined the
PMS data from the ONC and NGC 2264 by first transforming the ONC to
NGC 2264 age, using the results shown in the upper left panel, and
then combining these two clusters. Obviously there is not much
difference between doing this and transforming the individual
clusters.

Our conclusion from this exercise is that it is possible to
adequately model, at least in a statistical sense, the evolution of
all of the j distributions in terms of a simple scaling of a fraction
of the already slowly rotating population. Quantitatively, the
fraction affected is 40-50\% and the scale factors required are given
in Table \ref{model} and are quite substantial. How consistent are
these numbers, which are derived entirely from an empirical
assessment of the data, with the predictions of disk-locking theory?
As noted above, there is no quantitative disk-locking theory with
which to compare due to the theoretical difficulties mentioned above,
so we take the simple, first-order assumption that rotation period
remains constant during the disk-locked phase. Assuming a starting
radius of 2.09 R$_\odot$, appropriate to the ONC, one can estimate
the radius of the stars at the time the disk-locking must cease,
again assuming that the rotation period remains fixed. These radii
are given in Table \ref{model}. In the case of the comparison between
the ONC and NGC 2264 it is not necessary, of course, that the
disk-locked phase has, indeed, ceased. The radius given is simply the
radius to which the NGC 2264 stars must have contracted with constant
period to match the j distributions. 

Looking first at NGC 2264, we see that the derived value of R=1.8
R$_\odot$, based simply on the rotational period distributions and
the simplest possible assumptions consistent with a disk-locking
interpretation, is remarkably close to the median value adopted for
the cluster (R=1.7 R$_\odot$) from measurements of the luminosity and
effective temperatures of the stars. We take this excellent agreement
to be an indication that, to first order, the idea of disk-locking
(for 40-50\% of the stars) provides a good way of describing the
evolution of rotation from ONC age to NGC 2264 age. Going further, we
can ask how this might continue to the MS clusters. Here we find that
the typical radius at which disk-locking must end is about 1.2 
R$_\odot$, again under the assumption of a constant rotation period.
If disk-locking continued beyond that point, the stars in the MS
clusters would rotate too slowly to have evolved in this way from the
PMS distributions. It may be seen on the right hand panel of Fig.
\ref{hr_diagram} that the ``release point" of $\sim$1.2 R$_\odot$
happens to correspond roughly with the end of the Hayashi phase for
the more massive stars in our sample. The time to contract to such a
radius can, therefore, be estimated using the fact that R $\alpha$
t$^{-1/3}$ during the Hayashi phase and this leads to an estimate for
disk-locking times of about 5-6 My (see Table \ref{data}). To
summarize, a quantitative evaluation of the evolution of j with time
indicates that simple transformations of the data consistent with the
first-order ideas of disk-locking theory provide a wholly adequate
description of the data. The time scale required for the process of, at most, 5-6 My, is in good agreement with estimates of disk lifetimes based on near-infrared studies \citep{hll01}

\subsection{Selection Bias and Caveats}

It has been assumed that, in all clusters, the set of stars with
detected rotation periods is a representative sample in terms of
their rotation properties of the cluster as a whole. There are two
ways in which this assumption could be (and probably is, to some
extent) wrong. First, in the PMS clusters there is a likely bias
against finding rotation periods for slow rotators for the following
reason. Slow rotators are statistically more likely to be actively
accreting (i.e. Classical) TTS. The irregular variability that
accompanies accretion makes it more difficult to detect rotational
signals in the light curves of such objects. \citet{chw04}, for
example, have recently discussed this issue in some detail. Hence,
the j-distributions of the ONC (and NGC 2264) may have a bias against
slow rotators. In the ONC, where this comparison has been made, we
have found no significant difference between the v sin i
distributions of stars with and without rotation periods discovered
by spot modulation \citep{rhm01}. However, this study is not entirely
definitive because of the limited sample. A more detailed analysis by
\citet{hbm02} based on a number of considerations concluded that
there was a likely bias against slow rotators in the ONC sample but
probably only at about the 15\% level. 

In the MS sample there is also a likely bias against slow rotators,
but for a different reason. Clusters such as the Pleiades are so
spread out on the sky that they must be photometrically monitored on
an individual star basis, as opposed to including the entire cluster
on a single or few CCD images. The selection of which stars to
monitor for rotational variability may be biased if it is made with
reference to known v sin i measurements. J. Stauffer (private
communication) indicates that such a bias does indeed affect the
Pleiades sample used here since the observers (primarily he and C.
Prosser) preferentially selected known rapid rotators for study
assuming they would more likely yield measurable rotational periods
with the least investment of observational time. The extent of the
bias can be estimated by the fact that in the full sample of stars
with known v sin i, 49 out of 102 (48\%) have v sin i $<$ 10 km/s,
while among the stars with known rotation periods, only 14 out of 44
(32\%) fall in that category. This same bias probably does not affect
the other MS clusters to as great an extent (Stauffer, private
communication). 

Our conclusion is that both the PMS and MS samples are probably
biased to some extent against slow rotators, but that the degree of
biasing is probably only of the order of 15\%. This is a relatively
small effect that might act to increase somewhat the estimated
fractions of disk-locked stars if we had a more representative
rotational sample. In the future, it might be possible to evaluate
this effect more definitively by increasing the numbers of stars with
v sin i measurements and to lessen the effect by obtaining more
rotation periods for the slower rotators in the Pleiades. For now, it
is hard to see how this bias could affect our principal results. Only
in the case that we were missing a substantial number of very slow
rotators in the ONC would much would change on Fig.
\ref{j_hists_uncorrected}. Since PMS monitoring programs often extend
over several months they would have no difficulty detecting very slow
rotators if they existed so apparently, they do not. Therefore, just
to populate the slowly-rotating star bins of the current MS
distribution requires a good deal of loss of angular momentum from a
sizable fraction of stars. If there are even more slow rotators in
these MS clusters than is represented by the distributions shown
here, then disk-locking must be even more common than our current
analysis indicates. 

Finally, we should explicitly address the underlying assumption of
this analysis that the initial j distributions of the five clusters
considered were enough alike that the differences we measure today
reflect evolutionary effects and not initial conditions. The main
reason for such an assumption is that no progress can be made without
it. If some or all measured differences are assigned to initial
conditions then we can say nothing about evolution. If theory or
observations are someday able to establish that the Pleiades had a
much different j-distribution when it was 1 My old than the ONC has
today, our analysis and interpretation is obviously invalid. Given
the current state of the field we suspect this will not happen for a
long time, if ever. This paper shows that the current j distributions
can be understood in terms of an evolutionary sequence from a common
initial distribution represented by the ONC if one simply allows
angular momentum to be conserved for about half the sample and
disk-locking to affect the other half for about 5 My. At present, we
find no inconsistencies in the data that would seem to require that
we abandon the simplifying assumption of a common initial
j-distribution among the clusters included in this analysis. 

\subsection{Comparison with Results Obtained by Other Authors}

The question of angular momentum evolution of solar-like stars from
PMS to MS has been addressed frequently over the past few years by a
number of authors, as cited throughout this paper. To some extent
there has been disagreement on the following issues: 1) the bimodal
nature of the period (or j) distribution in the ONC, 2) the bimodal
nature of the period (or j) distribution in NGC 2264, 3) whether
there is evidence for spin-up of stars between the ONC and NGC 2264
ages, and 4) whether slowly rotating stars in the ONC or NGC 2264
actually have active accretion disks. See the review by \citet{m03}
for a concise summary of much of the debate. On the other hand, there is substantial agreement on the main features of the rotational evolution of solar-like stars as summarized in the Abstract of this paper and in Section V (Summary), which follows.

Here we would like to address areas of disagreement in the light of the new
results presented in this study and emphasize the agreement that exists on some major points. On the
bimodal nature of the period distribution in the ONC, we think it is
fair to say that the issue is entirely resolved now to everyone's
satisfaction. The source of the controversy was that several studies
of ``Orion" did not find a clearly bimodal period distribution
similar to what was first reported by \citet{ah92} and \citet{ch96}.
Two of these studies \citep{r01, chs01} were, in fact, not focussed
on the ONC but on the greater Orion association. The third
\citep{smm99} did not cover the same mass range. As additional data
have accumulated from a variety of sources, the original result has
been strengthened. Fig. \ref{per_hist_onc} contains all of the
currently available data on rotation periods in the ONC for stars in
the relevant mass range. There is no question that the solar-like
stars in the ONC have a bimodal period distribution.

In retrospect, it is not surprising in the least, but indeed
expected, that the period distributions for other samples of stars
which were neither limited by mass nor by position on the sky should
show a different period distribution. We now know, for example, that
less massive stars in the ONC spin faster \citep{hbm01} so mixing
masses within a period distribution tends to wipe out structure such
as bimodality. This is probably the main reason why \citet{smm99} did not find a
bimodal period distribution. Their sample contained many more low
mass stars than the samples analyzed by \citet{ah92} and
\citet{ch96}. In fact, when one limits the \citet{smm99} sample to
stars in the mass range considered here, it is distinctly bimodal.
These points have been raised and expounded upon in several papers,
including most recently by \citet{hbm02}. Hopefully, it will now be
clearly recognized that the bimodal nature of the ONC is not actually
a controversial subject any longer.

It is also not controversial that the greater Orion association (the
``flanking fields") of \citet{r01} and the survey of \citet{chs01}
have a greater preponderance of rapidly rotating stars than the ONC
and do not show a clearly bimodal distribution. Our interpretation of
this fact is that these samples are likely to contain a good mixture
of older stars associated with earlier star forming episodes in the
greater Orion association. Being older, these stars would have had
more time to contract and spin up, just as about half of the NGC 2264
stars have spun up with respect to the ONC. In fact, we would argue
that the average age of this more heterogeneous (than the ONC)
population is probably close to NGC 2264's average age of 2 My, since
the rotation period distributions of what \citet{mrs04} call ``Orion"
and NGC 2264 are not significantly different according to them.    

We also
note that previous studies of the Orion region, as listed above, have
found the ONC to be the youngest portion of the cluster and that
estimating ages by radii is difficult. Furthermore, when
\citet{lmb04} attempted to divide their NGC 2264 sample into a
younger and older half by location on an HR diagram they found no
significant difference in the rotation properties. Taken at face
value, this would mean that older, more contracted stars do not spin
faster than their younger counterparts, contradicting the results of
this study. In fact, we believe, it is simply another indication of
the fact that determining ages of PMS stars by determining their
radii is fraught with difficulty. 

That brings us to the conclusions of \citet{rws04}, who found no evidence for spin-up of stars due to
contraction during the PMS phase but did conclude that 30-40\% of the stars on convective tracks in the relevant mass range could have been released by the time they are 1 My old. Apparently their PMS sample was a little too small at any given age to identify the subtle change in the broad j-distributions that occurs between 1 and 2 My. On the other hand, their main conclusion, based on comparing the PMS to the recently arrived MS clusters is not too different from ours. We simply find 10-30\% more stars populating the rapid rotator portion of the sample.  Further discussion of these points has been included in \citet{lmb04} and need not be pursued here. The distinction between 30-40\% of the stars conserving angular momentum and 50-60\% is probably small enough that we should be more impressed with the similarity of these numbers than their difference. The studies agree in pointing to disk-locking as the most likely source of the angular momentum drain.

One line of argument that is sometimes raised against disk-locking is
that it does not work in detail, i.e. that slowly rotating stars in
the PMS clusters do not appear preferentially to have disks. Because
this argument is often repeated we reiterate here that it is not
true. A statistically significant correlation between rotation and
various disk indicators such as near-infrared excess and H$\alpha$
emission strength has been shown to exist in both the ONC and in NGC
2264 \citep{hbm02,lmb04,ds05}. Apparently, some investigators feel
that these correlations should be tighter than they are to be
convincing, even though they have a high statistical significance.
Our opinion is that there is a lot of scatter introduced into the
relationships by the inherent variability of TTS, the difficulty of
detecting disks and the time scales for disk dissipation and
subsequent spin-up.  While rotation rates can be accurately measured
to 1\% and respond only slowly (i.e. on time scales of 10$^5 - 10^6$
y) to external influences, indications of the presence or absence of
disks are notoriously difficult to measure and some, including H$_alpha$ equivalent width, ultraviolet excess and even near-infrared excess, can vary on time
scales as short as hours or days \citep{hhg94, chs01}. In our opinion, it is that variability and
observational difficulties which make the scatter in relations
between rotation and disk properties so large. The fact that we can,
in spite of this scatter, find statistically significant correlations
between rotation and disk indicators, is hard to understand if there
were no physical link. Obviously, this is an area in which additional
monitoring and observations of selected stars should prove fruitful.
  
\section{Summary}

\begin{figure}
\plotone{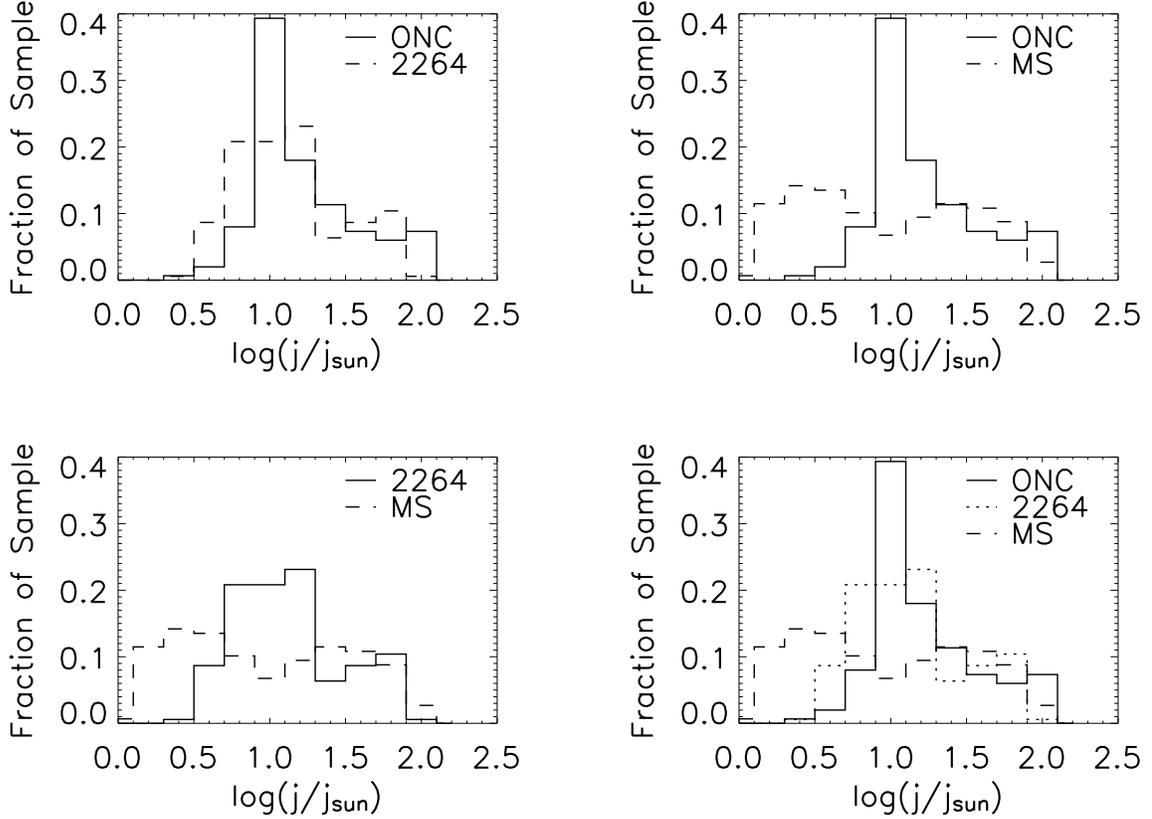}
\caption{The observed
j-dstributions of the ONC, NGC 2264, and the combined three other
clusters corrected for wind losses (labeled MS for ``main sequence").
It is clear that there is little change on the high-j side, implying
that rapidly rotating stars nearly conserve angular momentum as they
evolve from the PMS to the MS. However, there is a broadening of the
distribution on the low-j side which is noticeable in the comparison
of the 1 My old ONC with the 2 My old NGC 2264 clusters and becomes
quite dramatic when comparing the PMS and MS clusters. This indicates
that slowly rotating PMS stars must lose substantial additional
amounts (factor of 3 or more) of their surface angular momentum
during contraction to the MS. These effects are in agreement with
expectation based on disk-locking theory. }
\label{j_hists_uncorrected}
\end{figure}

\begin{figure}
\plotone{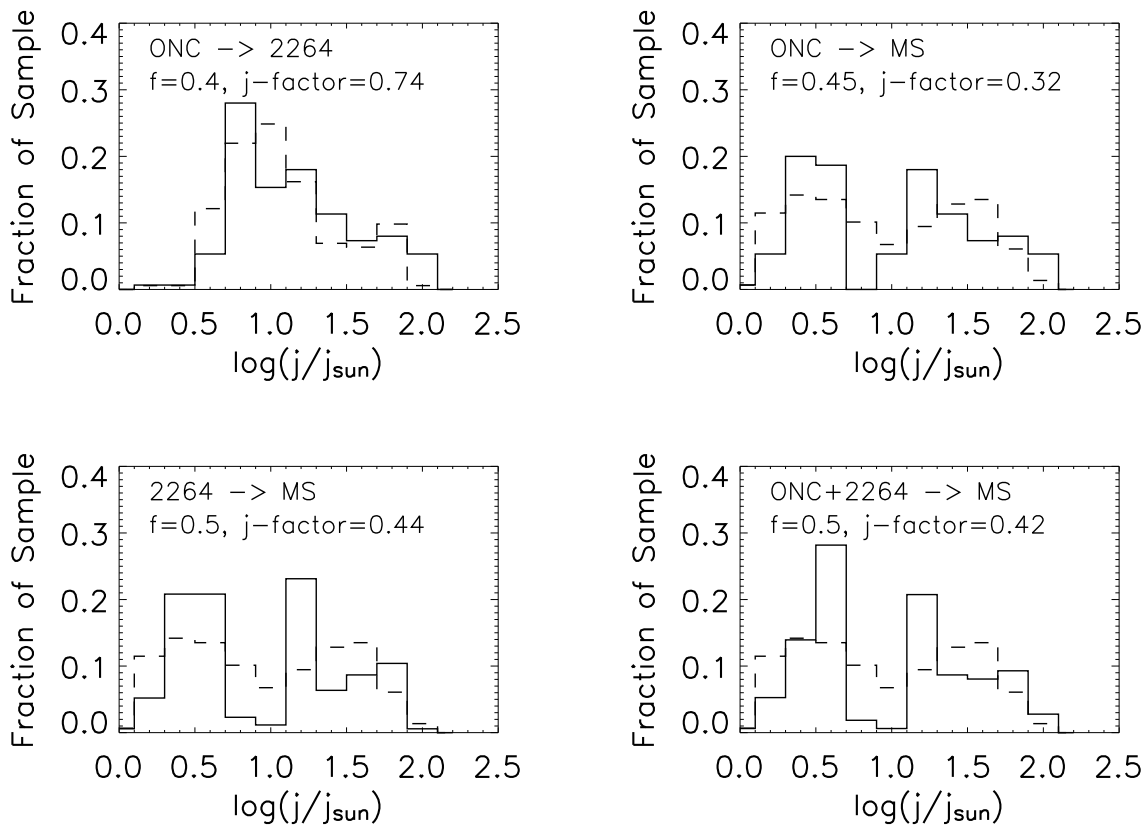}
\caption{The j
distributions of the ONC and NGC 2264 (corrected for 10\% wind
losses) have been adjusted (on the low angular momentum side) by the
j-factors indicated on each panel, which is applied to the fraction
(f) having the lowest j values. These are the simplest
transformations that give adequate fits to the data. Clearly, one
requires shifts by large factors applied to 40-50\% of the stars on
the slow rotating side of the distributions.}
\label{models}
\end{figure}

We have formed j distributions for stars of solar-like mass at three
different ages: nominally 1, 2 and 50 My. The distributions at all
times are broad and bimodal. The rapidly rotating side evolves with
only a small or negligible loss of angular momentum while the slowly
rotating side shows much greater angular momentum losses as expected
from disk-locking (see Fig. \ref{j_hists_uncorrected}). The data
indicate that a broad range of rotation rates is established within 1
My, presumably by magnetic interactions between the stars and their
accretion disks. The subsequent rotational evolution can be
characterized as spin-up during PMS contraction with conservation of
angular momentum (plus, perhaps, a small amount of angular momentum
loss through a stellar wind) for at least half of the stars. About
40-50\%, however, all of which are already slow rotators at 1 My,
must lose substantial additional amounts of angular momentum
($\sim$70\%) by the time they reach the MS. Furthermore, they must
lose a good deal of this angular momentum quickly. In the short time
interval ($\sim$1 My) between ONC and NGC 2264 ages, while most stars
spin up conserving angular momentum, 40-50\% do not. The observed
size of the angular momentum loss, the fact that it affects
preferentially the slowly rotating side of the j distribution, and
its rapid action over times as brief as 1 My, all support an
interpretation of disk-locking as the physical cause. We argue that
all of the currently available observational evidence on PMS rotation
and disks is consistent with this picture: one-half, or more, of
solar-like stars in clusters are no longer locked to their disks by 1
My while 40-50\% maintain such a locking for times of order 5 My.  These results are in reasonable agreement with what others have found for disk-locking times, percentage of stars affected and disk survival times based on infrared excess measurements \citep{hll01, tpt02, wsh04}.  

We note that our interpretation is based on two necessary assumptions
which can be tested by further observation. First, we assume that the
j-distributions derived from rotation periods are representative of
the full cluster j-distributions. If rotation period determinations
are significantly biased against slow rotators, as is likely at some
level, we may have underestimated the percentage of disk-locked
stars. Such tests as can be done at present suggest that the effect
is only at the $\sim$15\% level at most and will, therefore, not have
a major impact on our results. Second, we assume that the other four
clusters in our sample had j-distributions similar to that displayed
by the ONC when they were at a similar age. If that is not true, then
differences between their current j-distributions that we attribute
to evolution might, in fact, be due to differences in initial
conditions. The only way to check on this possibility is to increase
the number of clusters at all ages that have sufficient rotation
periods to define distributions. This will not be easy because
appropriate clusters are further away and will require extended
periods of observation on larger telescopes than have been used
heretofore. Finally, we note that our results apply only to stars of
solar-like mass (0.4 - 1.2 M$_\odot$). It would be interesting to
know if things were different for lower mass stars, but that will
require many more rotation periods for low mass stars in MS clusters,
an observationally challenging task.

\acknowledgments
 We thank John Stauffer for information and advice regarding the periods
and radii of the MS stars and for a critical reading of a first draft
of this manuscript which led to substantial revisions. We thank our
close and long-time collaborators Coryn Bailer-Jones, Markus Lamm,
Catrina Hamilton and Eric Williams for help with various aspects of
this study. We thank the referees, Sidney Wolff and Steve Strom for their constructive suggestions on the original submitted manuscript.We are indebted to the many Wesleyan students who, for
more than two decades, have manned the telescope to obtain data for
this project. Finally,
we thank NASA for their support of this work over the years through
its Origins of Solar 
Systems program, most recently grant NAG5-12502
to W.H.

\clearpage

\begin{deluxetable}{ccc}
\tabletypesize{\scriptsize}
\tablecaption{K-S Probabilities \label{data}}
\tablewidth{0pt}
\tablehead
{
\colhead{Cluster Comparison}   & 
\colhead{Sample}   &
\colhead{K-S Prob.}}

\startdata
& {\bf No Adjustments} & \\
 ONC and NGC 2264 & full distribution &
$1.6 \times 10^{-3}$  \\
 & rapid rotator side& 0.21  \\
 & slow rotator side&  $9.0 \times 10^{-7}$ \\
 ONC and MS & full distribution & $1.9 \times  10^{-11}$  \\
  & rapid rotator side& 0.48  \\
 & slow rotator side&  $9.1 \times  10^{-23}$ \\
 NGC 2264 and MS & full distribution & $1.6 \times  10^{-8}$  \\
  & rapid rotator side& 0.10  \\
 & slow rotator side&  $8.7 \times  10^{-17}$ \\
  & & \\
& {\bf Wind Losses\tablenotemark{a}} & \\
  ONC and MS &
full distribution & $2.3 \times  10^{-10}$  \\
  & rapid rotator side& 0.54  \\
 & slow rotator side&  $1.3 \times  10^{-20}$ \\
 NGC 2264 and MS & full distribution & $1.4 \times  10^{-7}$  \\
  & rapid rotator side& 0.018  \\
 & slow rotator side&  $7.1 \times  10^{-15}$ \\
  & & \\
& {\bf Disk Locking\tablenotemark{b}} & \\
  ONC and NGC 2264 & full distribution & $0.63$  \\
  ONC and MS &
full distribution & $0.25$  \\
   NGC 2264 and MS & full distribution & $0.47$  \\
  (ONC + 2264) and MS & full distribution & $0.25$  \\

 \enddata
\tablenotetext{a}{In this case all j values of the ONC
and NGC 2264 have been multiplied by the factor 0.9 (see text) to
simulate the effect of wind losses.}
 \tablenotetext{b}{In this case we multiply a fraction of the slow
side of the distribution by various factors to simulate the effect of
disk-locking acting on the slow rotators (see Fig,\ref{models} for
fractions and factors).}

\end{deluxetable}
 
 \begin{deluxetable}{ccccc}
\tabletypesize{\scriptsize}
\tablecaption{ Properties of Acceptable Disk-Locking
Models\label{model}}
\tablewidth{0pt}
\tablehead
{
\colhead{Cluster
Comparison}   & 
\colhead{f\tablenotemark{a}}	&
\colhead{j-factor\tablenotemark{b}} &
 \colhead{R\tablenotemark{c} (solar radii)} &
 \colhead{DLT\tablenotemark{d} (My)}
}
\startdata
  ONC and NGC 2264 & 0.4 & 0.74 & 1.80 &  \\
  ONC and MS & 0.45 &
0.32 & 1.18 & 5.5 \\
   NGC 2264 and MS & 0.5 & 0.44 &  1.19 & 5.4 \\
  (ONC + 2264) and MS & 0.5 & 0.42 & 1.17 &  5.8 \\

 \enddata
\tablenotetext{a}{The fraction (f) of the whole sample
which must be disk-locked.}
 \tablenotetext{b}{The factor by which the j-values of that fraction
are scaled.}
 \tablenotetext{c}{The radius to which the stars have contracted
while remaining disk-locked, assuming a constant period and starting
radius of 2.09 R$_\odot$.}
 \tablenotetext{d}{The (disk locking) time (DLT) required for
contraction to the radius in column 4, assuming that R $\alpha$
t$^{-1/3}$, as is appropriate to Hayashi-phase contraction.}
\end{deluxetable}

\clearpage

\clearpage

\clearpage

\clearpage

\clearpage

\clearpage

\clearpage

\clearpage

\clearpage

\clearpage

\clearpage

\clearpage

\clearpage

\clearpage

\clearpage

\clearpage

\clearpage


\clearpage

\end{document}